\documentclass[a4paper,11pt]{article}
\pdfoutput=1 
\usepackage{jcappub} 
\usepackage[T1]{fontenc} 
\usepackage{subfigure}
\newcommand{\Planck}{{\textit{Planck }}}

\newcommand{\CAMB}{{\tt{CAMB }}}
\newcommand{\MGCAMB}{{\tt{MGCAMB }}}

\newcommand{\expec}[1]{\left\langle #1\right\rangle}
\newcommand{\half}[0]{\frac{1}{2}}

\newcommand{\ld}[0]{\mathcal{L}}

\newcommand{\dd}[0]{\textrm{d}}
\newcommand{\defn}[0]{\equiv}

\newcommand{\qsubrm}[2]{{#1}_{\scriptsize{\textrm{#2}}}}
\newcommand{\qsuprm}[2]{{#1}^{\scriptsize{\textrm{#2}}}}
\newcommand{\qsubprm}[3]{{#1}^{\scriptsize{\textrm{#2}}}_{\scriptsize{\textrm{#3}}}}

\newcommand{\tis}[0]{ {\theta}^{\scriptscriptstyle\rm{S}}}
\newcommand{\tisdot}[0]{ {\dot{\theta}}^{\scriptscriptstyle\rm{S}}}
\newcommand{\pis}[0]{ {\Pi}^{\scriptscriptstyle\rm{S}}}

\newcommand{\rbm}[1]{{\bf{#1}}}

\newcommand{\kin}[0]{{\mathcal{X}}}
\newcommand{\hct}[0]{\mathcal{H}}
\renewcommand{\figurename}{Figure}

\def\be{\begin{equation}}
\def\ee{\end{equation}}
\def\bea{\begin{eqnarray}}
\def\eea{\end{eqnarray}}
\def\bse{\begin{subequations}}
\def\ese{\end{subequations}}

\newcommand{\fref}[1]{{Fig.~\ref{#1}}}

\let\oldsqrt\sqrt
\def\sqrt{\mathpalette\DHLhksqrt}
\def\DHLhksqrt#1#2{%
\setbox0=\hbox{$#1\oldsqrt{#2\,}$}\dimen0=\ht0
\advance\dimen0-0.2\ht0
\setbox2=\hbox{\vrule height\ht0 depth -\dimen0}%
{\box0\lower0.4pt\box2}}

\usepackage{soul}
\def\mnras{MNRAS}

\title{Constraining dark sector perturbations I: cosmic shear and CMB lensing}

\author[a]{Richard A. Battye,}
\author[b]{Adam Moss,}
\author[c]{and Jonathan A. Pearson}

\affiliation[a]{Jodrell Bank Centre for Astrophysics, School of Physics and Astronomy, The University of Manchester, Manchester, M13 9PL, U.K.}
\affiliation[b]{School of Physics \& Astronomy, University of Nottingham, Nottingham, NG7 2RD, U.K.}
\affiliation[c]{Centre for Particle Theory, Department of Mathematical Sciences, Durham University, South Road, Durham, DH1 3LE, U.K.}

\emailAdd{richard.battye@manchester.ac.uk}
\emailAdd{adam.moss@nottingham.ac.uk}
\emailAdd{jonathan.pearson@durham.ac.uk}

\usepackage{tikz}
\usetikzlibrary{shapes,arrows}

\abstract{We present current and future constraints on equations of state for dark sector perturbations. The equations of state considered are those corresponding to a generalized scalar field model and time-diffeomorphism invariant $\ld(g)$ theories that are equivalent to  models of a relativistic elastic medium and also Lorentz violating massive gravity. We develop a theoretical understanding of the observable impact of these models. In order to constrain these models we use CMB temperature data from {\it Planck}, BAO measurements, CMB lensing data from {\it Planck} and the South Pole Telescope, and weak galaxy lensing data from {CFHTLenS}. We find non-trivial exclusions on the range of parameters, although the data remains compatible with $w=-1$.  We gauge how future experiments will help to constrain the parameters. This is done via a likelihood analysis for CMB experiments such as CoRE and PRISM, and tomographic galaxy weak lensing surveys, focussing in on the potential discriminatory power of \textit{Euclid} on mildly non-linear scales.}

\begin{document}
\maketitle
\flushbottom

\section{Introduction}

There has been a recent explosion in the number of dark energy \cite{Copeland:2006wr} and modified gravity \cite{Clifton:2011jh} theories constructed in an attempt to describe the apparent observation of cosmic acceleration \cite{Perlmutter:1998np, Riess:1998cb}. It is now standard practice to include some ``dark contribution'' into the material content of the universe during the analysis of, e.g., observations of the temperature fluctuations in the Cosmic Microwave Background (CMB) \cite{Ade:2013zuv} and gravitational lensing of the CMB \cite{vanEngelen:2012va, 2013arXiv1303.5077P}.

Observations of the CMB \cite{2012arXiv1212.5225B, vanEngelen:2012va, 2013arXiv1303.5077P, 2013arXiv1303.5062P, 2013arXiv1303.5075P} and galaxy weak lensing \cite{2012MNRAS.427..146H, 2013MNRAS.430.2200K} are now mature enough to be able to use the current data   to constrain some of the allowed properties of the dark sector (by which we mean dark energy and/or modified gravity, rather than dark matter). Recent dark energy constraints have  focused on constraining the constant equation of state parameter $w$ and ``quintessence'' sound speed $\qsubrm{c}{s}^2$. Since these types of dark energy models do not cluster significantly, the  constraints are rather weak \cite{PhysRevD.69.083503, Ballesteros:2010ks, dePutter:2010vy, Salvatelli:2013wra, Xia:2013dea}. There are a small number of constraints on a particular parameterization of modified gravity since the data release from \textit{Planck}  \cite{Marchini:2013oya, Hu:2013aqa}.
Prospective CMB experiments such as the Polarized Radiation Imaging and Spectroscopy Mission (PRISM)  \cite{Andre:2013afa} and  Cosmic Origins Explorer (CoRE) \cite{Bouchet:2011ck}, as well as future tomographic galaxy weak lensing experiments like the Dark Energy Survey (DES) \cite{Abbott:2005bi}, {\it Euclid} \cite{Laureijs:2011mu}, and Large Synoptic Survey Telescope (LSST) \cite{Abate:2012za}, will considerably improve the constraining power. However, for any of these experiments to be used to their full potential, some framework for the evolution of possible perturbations in the dark sector is vital.

A useful framework for the dark sector at the level of the background already exists, and works very well: the dark contribution is   modelled by the dark energy equation of state parameter $w = P/\rho$ (where $P$ and $\rho$ are the pressure and energy density of the dark fluid). However, as soon as observational data   are used which are sensitive to the clustering properties of the dark sector, some new framework needs to be employed. Such observations are those that are effected by the integrated Sachs-Wolfe effect and gravitational lensing. There is   an abundance of parameterizations for perturbations of the dark sector in the literature \cite{PhysRevD.81.083534, PhysRevD.81.104023, Baker:2011jy, Zuntz:2011aq, Baker:2012zs, Mueller:2012kb, Kunz:2012aw, Silvestri:2013ne, Motta:2013cwa}, but a clear problem is emerging. A parameterization may well be constructed which covers the entire space of theories, but it will undoubtedly come with a huge number of free functions which experimental   data   will be unable to meaningfully constrain. Similarly, a very simple parameterization could be constructed which has a very small number of free parameters, but it may be too restrictive and will not probe wide enough ranges of theory space to yield theoretically meaningful results. And so:  a balance which must be struck between theoretical generality and observational feasibility. 

In this paper we will review and constrain the freedom in our \textit{equations of state for perturbations} formalism. The parameterization contains all theories with a given field content and symmetries, and comes with a very small number of free parameters. This paper   is complementary to  \cite{constraints_II}, in which measurements of ISW and CMB lensing tomography are used to constrain the equations of state for perturbations.

The layout of this paper is as follows. In section \ref{section:review} we explain and review the equations of state for perturbations formalism, as well as providing details of the models we   consider. In section \ref{sec:density} we give some analytic evidence for the difference in the lensing potentials for classes of models with and without anisotropic stress: our fluids approach makes this particularly transparent. In section \ref{section:obsquants} we review the observational spectra which are used to obtain the data constraints presented in section \ref{section:presconstr}. Section \ref{section:futureconstr} describes the ability of future experiments at constraining the freedom in the models, and concluding remarks are saved for section \ref{section:conclusions}.

\section{Review of the formalism}
\label{section:review}
\subsection{Basic ideas}
At its heart, the \textit{equations of state for perturbations} formalism was designed to allow meaningful and model independent statements about the allowed properties of the dark sector to be extracted from observations. We will   briefly review the formalism introduced and developed in \cite{Battye:2012eu, Pearson:2012kb, BattyePearson_connections, PearsonBattye:eos, Bloomfield:2013cyf, Battye:2013ida}.

The gravitational field equations for the class of theories we will study can be written as
\bea
{G^{\mu}}_{\nu} = 8 \pi G \left[{T^{\mu}}_{\nu}+{U^{\mu}}_{\nu}\right],
\eea
where ${T^{\mu}}_{\nu}$ is the   energy-momentum tensor of all known matter fields, and ${U^{\mu}}_{\nu}$ is the  dark energy-momentum tensor which contains all contributions to the gravitational field equations due to the dark sector theory.  This class of theories includes genuine dark energy models and also modified gravity models where ${U^{\mu}}_{\nu}$ includes, for example, higher derivatives of the metric. We will only study the minimally coupled dark sector here: this means that ${T^{\mu}}_{\nu}$ and ${U^{\mu}}_{\nu}$ are separately conserved. This restriction can, in principle, be relaxed.

For a background which is homogeneous and isotropic, ${U^{\mu}}_{\nu}$ can be specified in terms of two functions: the density, $\rho = \rho(t)$, and pressure, $P = P(t)$, of the dark sector fluid. The evolution of $\rho$ is constrained by the conservation equation $\nabla_{\mu}{U^{\mu}}_{\nu}=0$, giving $\dot{\rho} = - 3 \hct(\rho+P)$ and hence the specification of an equation of state parameter, $w=P/\rho$, defines the evolution of the background.

The components of the perturbed dark energy-momentum tensor are parameterized as
\bea
\delta {U^{\mu}}_{\nu} = \delta\rho u^{\mu}u_{\nu} + 2(\rho+P)v^{(\mu}u_{\nu)} + \delta P{\gamma^{\mu}}_{\nu} + P{\Pi^{\mu}}_{\nu},
\eea
where $\delta\rho, v^{\mu}, \delta P$ and ${\Pi^{\mu}}_{\nu}$ are the perturbed fluid variables of the dark sector theory;  we use round brackets to denote symmetrization of the enclosed indices, e.g., $A_{(\mu}B_{\nu)} = \tfrac{1}{2}\left(A_{\mu}B_{\nu} + A_{\nu}B_{\mu}\right)$. $u^{\mu}$ is a time-like vector and $\gamma^{\mu\nu}=u^{\mu}u^{\nu}+g^{\mu\nu}$ is the spatial metric, that is orthogonal to the time-like vector: $u^{\mu}\gamma_{\mu\nu}=0$.
The components of $\delta {U^{\mu}}_{\nu}$ are constrained by the perturbed fluid equations $\delta(\nabla_{\mu}{U^{\mu}}_{\nu})=0$, which,  for scalar perturbations in the synchronous gauge, are
\bse
\label{eq:sec:fluid-eqs-fourier}
\bea
\dot{\delta} &=& (1+w)\big( k^2\tis - \tfrac{1}{2}\dot{h}\big) - 3 \hct w\Gamma,\\
\tisdot &=& - \hct (1-3w)\tis - \tfrac{w}{1+w}\big(\delta   + \Gamma - \tfrac{2}{3} \pis\big),
\eea
\ese
where the perturbed pressure $\delta P$ has been packaged into the gauge invariant combination
\bea
w\Gamma \defn \left(  \frac{\delta P}{\delta\rho} - w\right)\delta.
\eea
Here, $\delta, \tis, \delta P$ and $\pis$ are the scalar parts of the perturbed dark sector fluid variables  (note that we take $\dot{w}=0$ for simplicity).  In principle, knowledge of two of these functions specified the others through the equations of motion. In (\ref{eq:sec:fluid-eqs-fourier}), $\Gamma$ and $\pis$ are the gauge invariant entropy and anisotropic stress perturbations respectively and this property means that it is sensible to use these functions  to specify the modified gravity/dark energy theory. The fluid equations (\ref{eq:sec:fluid-eqs-fourier}) close when the entropy and anisotropic stress contributions are written as
\bse
\label{eq:sec:eos-prototype}
\bea
w\Gamma &=&A_1 \delta + A_2 \tis + A_3 \dot{h} + \cdots,\\
w\pis &=& B_1 \delta + B_2 \tis + B_3 \eta + \cdots. 
\eea
\ese
That is, when $w\Gamma$ and $\pis$ are specified as linear functions of variables which are already evolved ($h$ and $\eta$ are the metric perturbations in the synchronous gauge, as defined in \cite{Ma:1994dv}); note that the $\{A_i, B_i\}$ are chosen so that $\Gamma$ and $\pis$ are gauge invariant. The entropy $w\Gamma$ and anisotropic stress $w\pis$ written in the form of (\ref{eq:sec:eos-prototype}) are the {\it equations of state for perturbations}.

The main idea behind our approach is to specify two ingredients: (i) the field content of the dark sector, and (ii) ask for a particular set of symmetries or principles to be respected and not to specify the actual Lagrangian.  After these two ingredients are laid down, we are able to obtain \textit{all the freedom} at the level of linearized perturbations,  since these two ingredients are  sufficient information for   obtaining a precise form  of the coefficients $\{A_i, B_i\}$ in the equations of state for perturbations (with prescribed scale dependence), and there will be nothing else left to specify to characterize the perturbations.  In the next section we will provide the explicit examples we will be constraining in this paper.  


Clearly, what we could do  is to write down the largest field content  and the smallest set of symmetries imaginable, but the problem is that this would create more free parameters than current (and near-future) cosmological data sets are able to constrain. This generalization  is not of practical interest, at least for the foreseeable future.

\subsection{Survey of models}

Here we will describe what classes of models each of our equations of state for perturbations describes, and the number of   associated free parameters. Our models have   actions given by
\bea
S = \int\dd^4x\sqrt{-g}\bigg[ \frac{R}{16\pi G} - 2 {\ld}{}({\rm{X}}) \bigg] + \qsubrm{S}{m}[g_{\mu\nu};\Psi_i].
\eea
 As described above, each model is just a statement of the field content ``X'' in the dark sector Lagrangian $\ld$ and a set of symmetries (typically, but not exclusively, taken to be reparameterization invariance). We do not need to specify the functional form of the dark sector Lagrangian.

\subsubsection{Generalized scalar field models}
\label{sec:gsf} 
For all models with field content
\bea
\label{eq:sec:GSF-fieldcontent}
\ld = \ld(\phi, \partial_{\mu}\phi, \partial_{\mu}\partial_{\nu}\phi, g_{\mu\nu}, \partial_{\alpha } g_{\mu\nu}),
\eea
which are  reparameterization invariant, have second order field equations, and are linear in $\partial_{\alpha}g_{\mu\nu}$,   the equation of state for perturbations is \cite{PearsonBattye:eos}
\bea
\label{eq:sec:eos_geneeral}
w\Gamma &=& (\alpha - w)\bigg[ \delta - 3\beta_1 \hct(1+w)\theta^{\rm S} - \frac{3\beta_2\hct(1+w)}{2k^2 - 6(\dot{\hct} - \hct^2)}\dot{h} + \frac{3(1-\beta_2-\beta_1)\hct(1+w)}{6\ddot{\hct} - 18\hct\dot{\hct} + 6\hct^3 + 2 k^2\hct}\ddot{h}\bigg],\nonumber\\
\eea
and $\pis =0$. We call this class of theories the generalized scalar field (GSF) theories. There are just three free dimensionless functions, $\mathcal{F} \defn \{\alpha, \beta_1, \beta_2\}$. $\alpha$ and $\beta_2$ only depend  on time, but in general $\beta_1$ will have scale dependence given by $\beta_1 = \beta_1^{(0)}(t) + \beta_1^{(2)}(t)k^2$ as explained in \cite{PearsonBattye:eos}. In models where $\beta_1^{(2)}\ne 0$ such theories will diverge as $k\rightarrow\infty$ and, hence, we set $\beta_1^{(2)} \equiv 0$, and we are left with three free dimensionless functions of time. 

 It is important to realize that \textit{every} theory in this class is covered by   these three functions. In particular, the kinetic gravity braiding \cite{Deffayet:2010qz, Pujolas:2011he} theories are included, as are all cubic galileon theories \cite{Barreira:2013eea}.

As a subset of these theories, consider the dark sector Lagrangian with field content 
\bea
\label{eq:sec:SSF-fc}
\ld = \ld (\phi, \partial_{\mu}\phi, g_{\mu\nu}),
\eea
which  are also reparameterization invariant. The gauge invariant equations of state for perturbations are \cite{Weller:2003hw, PhysRevD.69.083503}
\bea
\label{eq:eos:sfde}
w\Gamma = (\alpha - w)\big[ \delta - 3  \hct(1+w)\theta^{\rm S}\big],\qquad \pis=0.
\eea
There is a single parameter, $\alpha$, which can be interpreted as a sound speed for sub-horizon modes.  It is clear that this simple scalar field model (\ref{eq:eos:sfde}) is recovered from the general case (\ref{eq:sec:eos_geneeral}) when $\beta_1 =1, \beta_2 = 0$. The equation of state for perturbations (\ref{eq:eos:sfde}) contains all minimally coupled $k$-essence theories whose Lagrangian is  $\ld = \ld (\phi, \kin)$, where $\kin \defn -\half(\partial\phi)^2$; the important point to realise is that by dialing the value of $\alpha$, one dials through \textit{all possible}  $\ld = \ld (\phi, \kin)$ theories. 

 If a functional form of the Lagrangian is known, then $\alpha = (1+2\kin\ld_{,\kin\kin}/\ld_{,\kin})^{-1}$. When $\alpha=1$, (\ref{eq:eos:sfde}) describes perturbations of minimally coupled quintessence, and when $\alpha=w$ the adiabatic perturbations of shift-symmetric $k$-essence.   

The theory whose Lagrangian is given by
\bea
\ld(\phi, \kin) = \bigg( -\kin + \frac{\kin^2}{M^4}\bigg)V(\phi),
\label{kess1}
\eea
where $M$ is a mass scale and $V$ an arbitrary function of $\phi$, has $\alpha$ related to $w$ via
\bea
\alpha = \frac{1+w}{5-3w}.
\label{kess2}
\eea

\subsubsection{Time diffeomorphism invariant $\ld(g)$ theories}
This class of models is rather distinct from  the other classes of models we will discuss. These  are constructed from a dark sector theory whose field content is
\bea
\label{eq:sec:TDI-fc}
\ld = \ld(g_{\mu\nu}),
\eea
and is constrained to be invariant under time diffeomorphisms, but not space diffeomorphisms. This leaves a residual vector degree of freedom and a massive graviton. We call this class of theories the time diffeomorphism invariant (TDI) $\ld(g)$ theories. These TDI $\ld(g)$ theories have been studied in the literature, under the name of \textit{elastic dark energy} \cite{PhysRevD.60.043505, PhysRevD.76.023005, Sitwell:2013kza}; the theory is equivalent to a particular brand of Lorentz violating massive gravities \cite{BattyePearson_connections}.

The equations of state for perturbations of the TDI $\ld(g)$ models are
\bea
\label{eq:eos:ede}
w\Gamma = 0,\qquad w\pis = \tfrac{3}{2}(w-\qsubrm{c}{s}^2 )\big[ \delta - 3(1+w)\eta\big],
\eea
where $\eta$ is the synchronous gauge metric perturbation.  There is a single parameter, $\qsubrm{c}{s}^2$, which is the sound speed of the medium. Notice that the medium is adiabatic (i.e., the entropy perturbation vanishes), but has an anisotropic stress. We should note that this theory represents one of the simplest possible ways to modify General Relativity.

\subsection{Summary}
Once  the perturbed fluid equations (\ref{eq:sec:fluid-eqs-fourier}) are furnished with one of the equations of state for perturbations, i.e., (\ref{eq:sec:eos_geneeral}) or (\ref{eq:eos:ede}), the system of perturbation equations  closes. In \fref{fig:shem_eos} we summarize the field contents, symmetries, and the associated number of free functions.


\tikzstyle{block} = [  text centered, rounded corners, minimum height=3em]
\tikzstyle{line} = [draw, -latex']
\tikzstyle{cloud} = [ minimum height=2em]
\begin{figure}[!t]
\begin{centering}
\begin{tikzpicture}[node distance = 2cm, auto]
    \node [block] (fullfield) {$\ld = \ld\left(\phi, \partial_{\mu}\phi, \partial_{\mu}\partial_{\nu}\phi, g_{\mu\nu}, \partial_{\alpha}g_{\mu\nu}\right)$};
    \node [block, right of=fullfield, node distance=7cm] (mainmodel_1) {$\ld = \ld\left(\phi, \partial_{\mu}\phi,g_{\mu\nu} \right)$};
    \node [block, right of=mainmodel_1, node distance=4.5cm] (mainmodel_2) {$\ld = \ld\left(g_{\mu\nu}\right)$};
    \node [cloud, below of=fullfield] (mp1) {$\left\{\alpha, \beta_1, \beta_2\right\}$};
        \node [cloud, below of=mainmodel_1] (mp2) {$\{\alpha \}$};
         \node [cloud, below of=mainmodel_2] (mp3) {$\left\{\qsubrm{c}{s}^2 \right\}$};
	\draw [line] (fullfield) -- node{restrict}(mainmodel_1);
	\draw [line] (mainmodel_1) --  node{restrict}(mainmodel_2);
	\draw [line] (fullfield) -- node[anchor=east] {diff.inv}(mp1);
	\draw [line] (mainmodel_1) -- node[anchor=east] {diff.inv}(mp2);	
	\draw [line] (mainmodel_2) -- node[anchor=east] {TDI}(mp3);		
	\draw [line] (mp1) -- node {$\beta_1 = 1, \beta_2 = 0$} (mp2);			
\end{tikzpicture}
\caption{Schematic depiction of how the field content and symmetries of the models we study are related. The field content on the far left (\ref{eq:sec:GSF-fieldcontent}) is imposed with reparameterization invariance (and second order field equations, as well as being linear in $\partial_{\alpha}g_{\mu\nu}$ for simplicity) to yield the three parameters $\{\alpha, \beta_1, \beta_2\}$ in the equation of state for perturbations (\ref{eq:sec:eos_geneeral}). Restricting the field content to (\ref{eq:sec:SSF-fc}) and then imposing reparameterization invariance leaves a single parameter, $\alpha$, in the equation of state for perturbations (\ref{eq:eos:sfde}). Further restriction of the field content to  (\ref{eq:sec:TDI-fc}) and then imposing time diffeomorphism invariance leaves a single parameter, $\qsubrm{c}{s}^2$ in the equations of state for perturbations (\ref{eq:eos:ede}). }\label{fig:shem_eos}
\end{centering}
\end{figure}

In the rest of the paper we observationally constrain allowed values of the free parameters: $\qsubrm{c}{s}^2$ in the TDI $\ld(g_{\mu\nu})$ models, and $\{\alpha, \beta_1, \beta_2\}$ in the     GSF models. In this paper we  restrict these parameters, and $w$, to be constant. We are simply being realistic about the constraining power of the available data. 

This constraint of constancy calls into question what portion of the GSF theories are actually covered by taking $\{\alpha, \beta_1, \beta_2\}$ as constant. However, if we are able to observationally constrain the characteristic size of these parameters, particularly if one of them has to be small, then we have in-fact used data to learn something about the allowed properties of the dark sector theory. By measuring the constant values of these parameters we are measuring something which is characteristic to the evolution of perturbations of theories of a given class.



\section{Evolution of the density contrast}
\label{sec:density}
We have cast our parameterization in terms of hydrodynamic quantities, and this can be used to obtain a simple, physically intuitive understanding of the role of the parameters in the equations of state for perturbations, and why they take on certain values could potentially yield an observable effect. In this section we will   have a brief look at the evolution equations of the density contrast of the dark  energy fluid, paying special attention to the qualitative difference in the evolution equation and lensing potential for a fluid with non-zero entropy and anisotropic stress.

Combining the fluid equations (\ref{eq:sec:fluid-eqs-fourier}) yields a second order evolution equation for the dark energy density contrast,
\bea
\label{eq:sec:evol_dens-pert}
&&\ddot{\delta}+  \hct(1-3w)\dot{\delta} +k^2w\delta \nonumber\\
&&\qquad=-\big[k^2+3\hct^2(1-3w) + 3 \dot{\hct}\big]w\Gamma  -3 \hct \dot{w\Gamma}  + \tfrac{2}{3}k^2w\Pi - \tfrac{1}{2} (1+w)\big[\ddot{h}+\hct(1-3w)\dot{h}\big] .\nonumber\\
\eea
On a Minkowski background (obtained   by setting $ a =1$), the evolution equation (\ref{eq:sec:evol_dens-pert}) becomes
\bea
\label{eq:sec:dens-ddot-flat-proto}
\ddot{\delta} + k^2w\delta = - k^2(w\Gamma - \tfrac{2}{3}w\Pi) - \tfrac{1}{2}(1+w)\ddot{h}.
\eea
We are interested in models with $w<-1/3$, so that the fluid accelerates the universe,  but without problematic gravitational instability which would be present for an adiabatic  perfect fluid: if $w\Gamma = w\Pi=0$, the density contrast only has exponential solutions when $w<0$, indicating the undesirable gravitational instability. However, in the GSF  (\ref{eq:sec:eos_geneeral}) and TDI $\ld(g)$ (\ref{eq:eos:ede}) models,  (\ref{eq:sec:dens-ddot-flat-proto}) becomes
\bse
\label{eq:sec:dens-ddot-flat-}
\bea
\label{eq:sec:dens-ddot-flat-sf}
\ddot{\delta} +\alpha k^2\delta &=& - \tfrac{1}{2}(1+w)\ddot{h},\\
\label{eq:sec:dens-ddot-flat-ede}
\ddot{\delta} + \qsubrm{c}{s}^2k^2\delta &=& - \tfrac{1}{2}(1+w)\big[\ddot{h} + 6k^2(w-\qsubrm{c}{s}^2)\eta\big]\,,
\eea
\ese
respectively.
Although both of these models are stable for $w<0$,  the physical origin is subtly different. In (\ref{eq:sec:dens-ddot-flat-sf}) perturbations are supported against exponential growth by the sound speed $\alpha$ of a fluid with entropy (non-adiabatic pressure) perturbations, whilst in (\ref{eq:sec:dens-ddot-flat-ede}) perturbations are supported due to the medium's   anisotropic stress. It is clear that perturbations in both cases are subluminal  and stable when  $0\leq \alpha \leq 1$ and $0\leq\qsubrm{c}{s}^2\leq 1$.
%
%

As we pointed out above, the physical origin of the protection against instability is different in the two cases: one was due to entropy, the other due to anisotropic stress. We can further elucidate this distinction by using the perturbed gravitational field equations to replace the metric fluctuations on the right-hand-sides of (\ref{eq:sec:dens-ddot-flat-}).   In the synchronous gauge, with $a=1$, the perturbed gravitational field  equations are given by
\bse
\label{grav-htc=0-cn}
\bea
-2k^2\eta &=& 8 \pi G\sum_{\rm i}\qsubrm{\rho}{i} \qsubrm{\delta}{i},\\
\dot{\eta} &=& - 4 \pi G\sum_{\rm i}(\qsubrm{\rho}{i} + \qsubrm{P}{i})\qsubrm{\theta}{i},\\
\ddot{h} - 2 k^2\eta &=& - 24 \pi G \sum_{\rm i}\qsubrm{\delta P}{i},\\
\ddot{h} + 6\ddot{\eta} - 2 k^2\eta &=& - 16 \pi G\sum_{\rm i}\qsubrm{P}{i} \qsubrm{\Pi}{i},
\eea
\ese
where the index ``i'' runs over all gravitating components: radiation, matter, and dark energy. After using (\ref{grav-htc=0-cn}) to replace all metric fluctuations in the combined perturbed fluid equations,  one obtains the following second order evolution equation for the density contrast of component ``i'':
\bea
\label{eq:sec:eom_flat_general}
\ddot{\delta}_{\rm i} + w_{\rm i}k^2\delta_{\rm i} &=& - k^2(w_{\rm i}\Gamma_{\rm i} - \tfrac{2}{3}w_{\rm i}\Pi_{\rm i})   +4\pi G (1+w_{\rm i}) \sum_{\rm j} \rho_{\rm j}[3w_{\rm j}\Gamma_{\rm j} + (1+3w_{\rm j})\delta_{\rm j}]. 
\eea
Suppose that there are  two components: ``dark energy'' and matter, $\rm i \in \{{\rm m}, {\rm de}\}$. The equations of state for perturbations for both the GSF and TDI $\ld(g)$ models yields the following sets of   coupled evolution equations
\begin{itemize}
\item GSF
\bse
\label{eom:gsf}
\bea
\label{gsf-m}
\ddot{\delta}_{\rm m} - 4 \pi G \rho_{\rm m}\delta_{\rm m}  =  4\pi G   (1+3\alpha) \rho_{\rm de}  \delta_{\rm de},
\eea
\bea
\label{gsf-de}
&&\ddot{\delta}_{\rm de} + \big[k^2\alpha- 4\pi G \rho_{\rm de}(1+w)(1+3 \alpha)\big]\delta_{\rm de} =   4 \pi G \rho_{\rm m}(1+w)\delta_{\rm m}.
\eea
\ese
\item TDI $\ld(g)$ 
\bse
\label{eom:TDI}
\bea
\label{lg-m}
\ddot{\delta}_{\rm m} - 4 \pi G \rho_{\rm m}\delta_{\rm m}  =  4\pi G   (1+3w)\rho_{\rm de}\delta_{\rm de},
\eea
\bea
\label{lg-de}
&&\ddot{\delta}_{\rm de} +\big[ \qsubrm{c}{s}^2k^2-4\pi G  \rho_{\rm de}(1+w)(1+6w-3\qsubrm{c}{s}^2 )\big]\delta_{\rm de}     =4 \pi G \rho_{\rm m}(1+w)[1+3(w-\qsubrm{c}{s}^2)]\delta_{\rm m}.\nonumber\\
\eea
\ese
\end{itemize}
Here we see that the evolution equations of the matter and dark energy density contrasts are qualitatively different in the two cases. To begin, notice that the source terms of the matter evolution equations (\ref{gsf-m}) and (\ref{lg-m}) are different: the GSF case is multiplied by $(1+3\alpha)$, but the TDI $\ld(g)$ case is multiplied by $(1+3w)$. Secondly, both the mass and source terms of the dark energy evolution equations (\ref{gsf-de}) and (\ref{lg-de}) are qualitatively different.

Using the standard relations between the synchronous and conformal Newtonian gauges \cite{Ma:1994dv} we can obtain an expression for the lensing potential for this mixture of matter and dark energy
\bea
\label{eq:sec:lensing-gen-flat}
K^2(\phi+\psi) = - 2 \left( \qsubrm{\delta}{m} + R \left[\qsubrm{\delta}{de} + \qsubrm{w}{de}\qsubrm{\Pi}{de}\right]\right),
\eea
where $R\defn \qsubrm{\rho}{de}/\qsubrm{\rho}{m}$ and $K \defn (4\pi G\qsubrm{\rho}{m})^{-1/2}k$. It is now clear that the  anisotropic stress  of dark energy has   a non-trivial effect on the (observable) lensing potential. In order to elucidate the distinction in the lensing potential for models with and without anisotropic stress, we will compute the normal modes of the equations of motion (\ref{eq:sec:eom_flat_general}) and express the lensing potential in terms of the growing and oscillating modes.

In the GSF and TDI $\ld(g)$ cases the equations of motion can be written as $ {\rbm{x}} ''+ \mathsf{A}\rbm{x} = 0$, where $\rbm{x} = \binom{\qsubrm{\delta}{m}}{\qsubrm{\delta}{de}}$ and the primes denote derivative with respect to rescaled time, $T \defn (4\pi G \qsubrm{\rho}{m})^{1/2}t$.  The matrix $\mathsf{A}$ for each of the equations of motion (\ref{eom:gsf}) and (\ref{eom:TDI}) is given by 
\bea
\label{eq:sec:BGSF}
\qsubrm{\mathsf{A}}{GSF} = \left( \begin{array}{cc} -1& - R(1+3\alpha)\\ -\epsilon & K^2\alpha -R(1+3\alpha)\epsilon \end{array}\right),
\eea
\bea
\qsubrm{\mathsf{A}}{TDI} = \left( \begin{array}{cc} -1& -R(1+3w)\\ - \left(1+3Q\right)\epsilon & \qsubrm{c}{s}^2K^2 - R(1+3[w+Q])\epsilon\end{array}\right)\,, 
\eea
respectively, where $\epsilon=1+w$ and $Q=w-\qsubrm{c}{s}^2$. The signs of the eigenvalues of the matrix $\mathsf{A}$ determines what type of solutions the matter and dark energy density contrasts posess: if $\lambda>0$, then solutions are oscillatory, and if $\lambda<0$ they are exponential and correspond to growing modes.  We will  construct the modes $\eta \defn \qsubrm{\delta}{m} + \mu\qsubrm{\delta}{de}$ that satisfy $\eta'' + \lambda \eta=0$.
 
One can verify that when $\alpha=0$ the GSF models have a zero eigenvalue for any value of $w$. For GSF models with $\alpha=0$ we form the combinations which correspond respectively   to the growing and zero eigenmodes,
\bse
\label{eq:sec:grow_0_GSF}
\bea
\qsubprm{\eta}{(GSF)}{grow} &=& \qsubrm{\delta}{m} + R\qsubrm{\delta}{de},\\
\qsubprm{\eta}{(GSF)}{zero} &=&  \qsubrm{\delta}{m} -\frac{\qsubrm{\delta}{de}}{\epsilon}.
\eea
\ese
The existence of a zero mode is not shared by   TDI models with $\qsubrm{c}{s}^2=0$, and as such it isn't possible to get such   aesthetically pleasing forms of the modes in the  TDI model for    general $w$. However, expanding for small $\epsilon$ reveals that there are     growing and oscillatory modes, which  are given by
\bse
\label{eq:modes-zerocs2}
\bea
\qsubprm{\eta}{(TDI)}{grow}=\qsubrm{\delta}{m}-\bigg[ 2 R-3 R (1+6 R) \epsilon+\mathcal{O}(\epsilon)^2\bigg] \qsubrm{\delta}{de},
\eea
\bea
\qsubprm{\eta}{(TDI)}{osc}&=&\qsubrm{\delta}{m}+\bigg[\frac{1}{2 \epsilon}+ \frac{3}{4} (1+6 R)  +\frac{9}{8}  \left(1-2 R-16 R^2\right)  \epsilon+\mathcal{O}(\epsilon)^2\bigg]\qsubrm{\delta}{de}. 
\eea
\ese  

Substituting the GSF modes (\ref{eq:sec:grow_0_GSF}) into the lensing potential (\ref{eq:sec:lensing-gen-flat}) reveals that
\bea
\label{eq:lensing-GSF-a0}
K^2(\phi+\psi) = - 2\qsubprm{\eta}{(GSF)}{grow},
\eea
and putting the TDI modes (\ref{eq:modes-zerocs2}) into the lensing potential (\ref{eq:sec:lensing-gen-flat})  yields
\bea
\label{eq:lensing-TDI-cs0}
 K^2 (\phi +\psi ) &=&\bigg[-2 +3R\left(4+ {3   }{K^{-2}}\right) \epsilon+\mathcal{O}(\epsilon)^2\bigg]\qsubprm{\eta}{(TDI)}{grow} +\bigg[ -12 R\epsilon+\mathcal{O}(\epsilon)^2\bigg]\qsubprm{\eta}{(TDI)}{osc}.\nonumber\\
\eea
There are some   important properties and distinctions between these expressions for the lensing potential. We first remind that both are given at $a=1$ and for $\alpha = \qsubrm{c}{s}^2=0$. The GSF lensing potential (\ref{eq:lensing-GSF-a0}) is valid for any value of $w$, and only depends on one mode (the growing mode) without any prefactors of the (rescaled) wavenumber $K^2$. The TDI lensing potential (\ref{eq:lensing-TDI-cs0}) includes both the growing and oscillating mode: the former also comes with a $K^{-2}$-dependence at $\mathcal{O}(\epsilon)$, and the latter only appears at $\mathcal{O}(\epsilon)$.

We now generalize to the case where $\alpha \neq 0$ and $\qsubrm{c}{s}\neq 0$. All expressions are sufficiently complicated that we exclusively expand to first order in $\epsilon$. The eigenvalues of $\mathsf{A}$ are
\bse
\bea
\qsubprm{\lambda}{(GSF)}{grow}&=&-\left[1+R\frac{ 1+3  \alpha   }{1+K^2 \alpha }\epsilon+\mathcal{O}(\epsilon)^2\right],\\
\qsubprm{\lambda}{(GSF)}{osc}&=&K^2 \alpha\left[ 1-R   \frac{ 1+3 \alpha    }{1+K^2 \alpha }\epsilon+\mathcal{O}(\epsilon)^2\right];
\eea
\ese
\bse
\bea
\qsubprm{\lambda}{(TDI)}{grow}&=&-\left[1+2R\frac{  2+3 \qsubrm{c}{s}^2  }{1+K^2 \qsubrm{c}{s}^2}\epsilon+\mathcal{O}(\epsilon)^2\right],\\
\qsubprm{\lambda}{(TDI)}{osc}&=&K^2 \qsubrm{c}{s}^2+\frac{ R}{1+K^2 \qsubrm{c}{s}^2} \bigg[{  9\left(1+\qsubrm{c}{s}^2\right) +(5+3\qsubrm{c}{s}^2) K^2 \qsubrm{c}{s}^2   }\bigg]\epsilon+\mathcal{O}(\epsilon)^2. 
\eea
\ese
The lensing potential in  the GSF case is given by
\bea
&&K^2(\phi +\psi ) \nonumber\\
&&=\left[-2  +2R \alpha\frac{  3-K^2       }{\left(1+K^2 \alpha \right)^2}\epsilon+\mathcal{O}(\epsilon)^2\right]\qsubprm{\eta}{(GSF)}{grow} +\left[-2R \alpha\frac{ 3-K^2     }{\left(1+K^2 \alpha \right)^2}\epsilon+\mathcal{O}(\epsilon)^2\right]\qsubprm{\eta}{(GSF)}{osc},\nonumber\\
\eea
and in the TDI case by
\bea
&&K^2(\phi +\psi ) \nonumber\\
&&= \bigg[-2+\frac{R}{\left(1+K^2 \qsubrm{c}{s}^2\right){}^2}\nonumber\\
&&\qquad\times\bigg( 9(1+\qsubrm{c}{s}^2)K^{-2}   +6(2+9\qsubrm{c}{s}^2 ) +c_s^2(4+15\qsubrm{c}{s}^2+9\qsubrm{c}{s}^3)K^2    \bigg)\epsilon+\mathcal{O}(\epsilon)^2\bigg]\qsubprm{\eta}{(TDI)}{grow}\nonumber\\
&&+\bigg[{   -2R  \frac{\left(2+3 \qsubrm{c}{s}^2\right) \left(3+K^2 \qsubrm{c}{s}^2\right)}{ \left(1+K^2 \qsubrm{c}{s}^2\right){}^2}  }{}\epsilon+\mathcal{O}(\epsilon)^2\bigg] \qsubprm{\eta}{(TDI)}{osc}. 
\eea
These are written in terms of the   modes $\{\qsubrm{\eta}{grow}, \qsubrm{\eta}{osc}\}$ which are easily computed, but we omit for brevity.

The  manipulations (with $a=1$)   we have presented highlights the qualitative difference in the evolution equations of the density contrasts in the two cases where the dark energy fluid has entropy and anisotropic stress. One of the important distinctions we draw out from this analysis  is the manner in which each mode influences the lensing potential, and the different scale dependences of the coefficients of the modes involved. This distinction will be fully vindicated when we evolve the equations of motion in the next section.


\section{Observational quantities}
\label{section:obsquants}
Let us, first, briefly review  the observational probes we intend to   use to constrain the equations of state for dark sector perturbations. We use a modified version of \CAMB \cite{Lewis:1999bs} to evolve the gravitational field equations with a dark sector fluid whose perturbations are parameterized via  equations of state. This provides the gravitational potentials which can be used to compute all observational quantities we will discuss.

We stress that the modification to \CAMB is minimal (of the order 10 lines of code): the equations of state involve quantities which are already evolved (or are trivially computable). This is in stark contrast to, e.g., the modifications to \CAMB that are required for the slip and effective gravitational coupling parameterizations in \MGCAMB \cite{Hojjati:2011ix}.
 
We compute the usual CMB observables: the CMB temperature and polarization angular power spectra,  as well as the CMB  lensing power spectrum, $C_{\ell}^{\phi \phi}$.  We also  compute the  galaxy lensing convergence power spectrum, $P_{ij}^{\kappa}(\ell)$, which in the absence of anisotropic stress can be written in terms of the matter power spectrum, $P(k,z)$. However, since our models include anisotropic stress, we use the more general form of the convergence power spectrum  
\bea
P_{ij }^{\kappa}(\ell) \approx  {2\pi^2\ell}  \int \frac{\dd \chi}{\chi} g_{i} \left(\chi \right)  g_{j} \left(\chi \right) P_{\Psi}  \left(\ell/ \chi, \chi \right)\,,
\eea
where we have made use of the Limber approximation in flat space, $\chi$ is the comoving distance, and $P_{\Psi}$ is the power spectrum of the Weyl potential, defined in terms of the Newtonian potentials $\phi$ and $\psi$ by $\Psi = \left(\psi + \phi \right)/2$.  The lensing efficiency is given by
\bea
g_i (\chi) = \int_{\chi}^{\infty} \dd \chi^{\prime} n_{i} \left( \chi^{\prime} \right) \frac{\chi^{\prime}-\chi}{\chi^{\prime}}\,,
\eea
where $n_{i} \left( \chi \right)$ is the radial distribution of source galaxies in bin $i$. In the case of no anisotropic stress, the convergence power spectrum can be written in the usual form 
\bea
\label{eq:sec:weak-lensing-ij-conv}
P_{ij}^{\kappa}(\ell) &=&\frac{9}{4}\qsubrm{\Omega}{m}^2 H_0^4\int_0^{\infty}\frac{g_i(\chi)g_j(\chi)}{a^2(\chi)}  P (\ell/\chi, \chi)\dd\chi\,. 
\eea
The convergence can also be written in terms of the correlation functions $\xi^{\pm}$ via
\bea
\xi_{i,j}^{\pm}(\theta) = \frac{1}{2\pi}\int \dd\ell\, \ell\, P_{ij}^{\kappa}(\ell) J_{0,4}(\ell \theta),
\eea
where $J_{0,4}$ are Bessel functions of the zeroth and fourth order respectively. 

\begin{figure}[!t]
      \begin{center}
\subfigure[\, TDI $\ld(g)$ models]{\includegraphics[scale=0.6,angle=0]{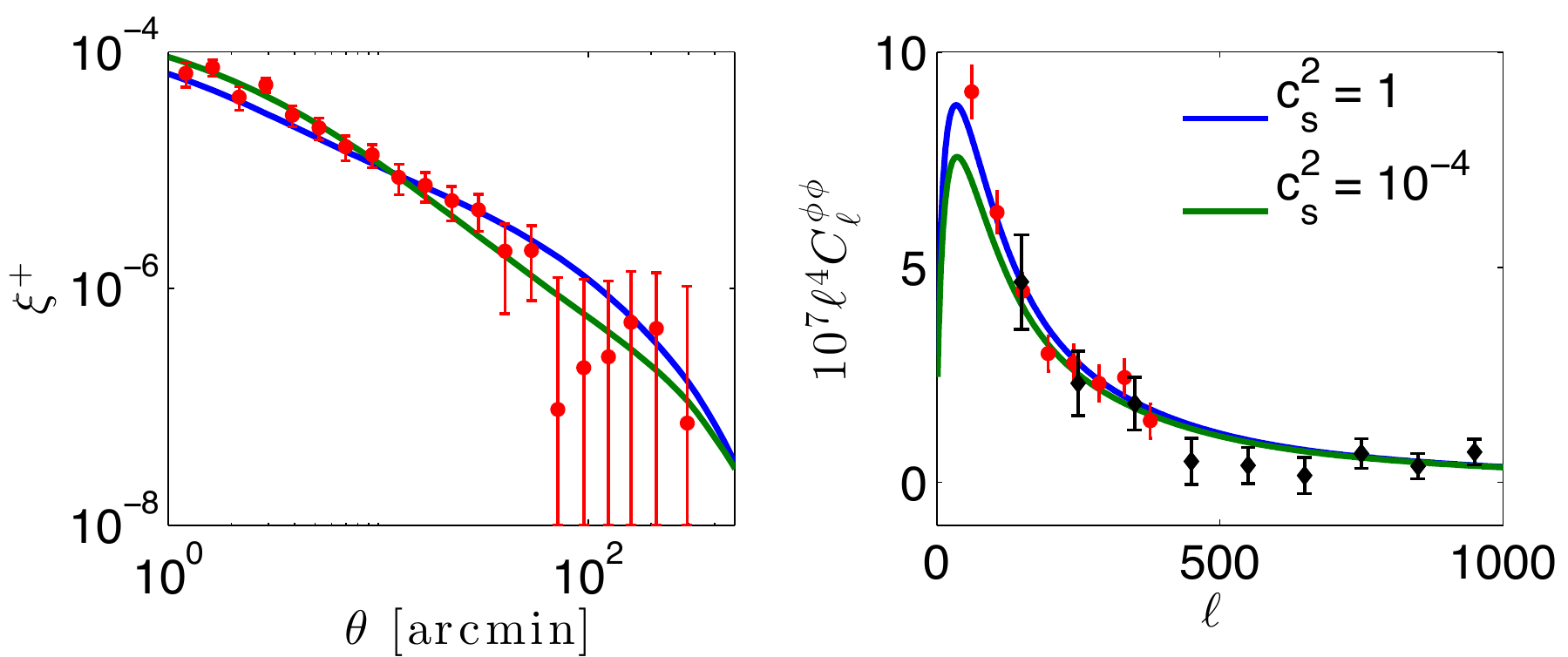}}
\subfigure[\, GSF models with $\alpha=0, \beta_2 = 1-\beta_1$]{\includegraphics[scale=0.6,angle=0]{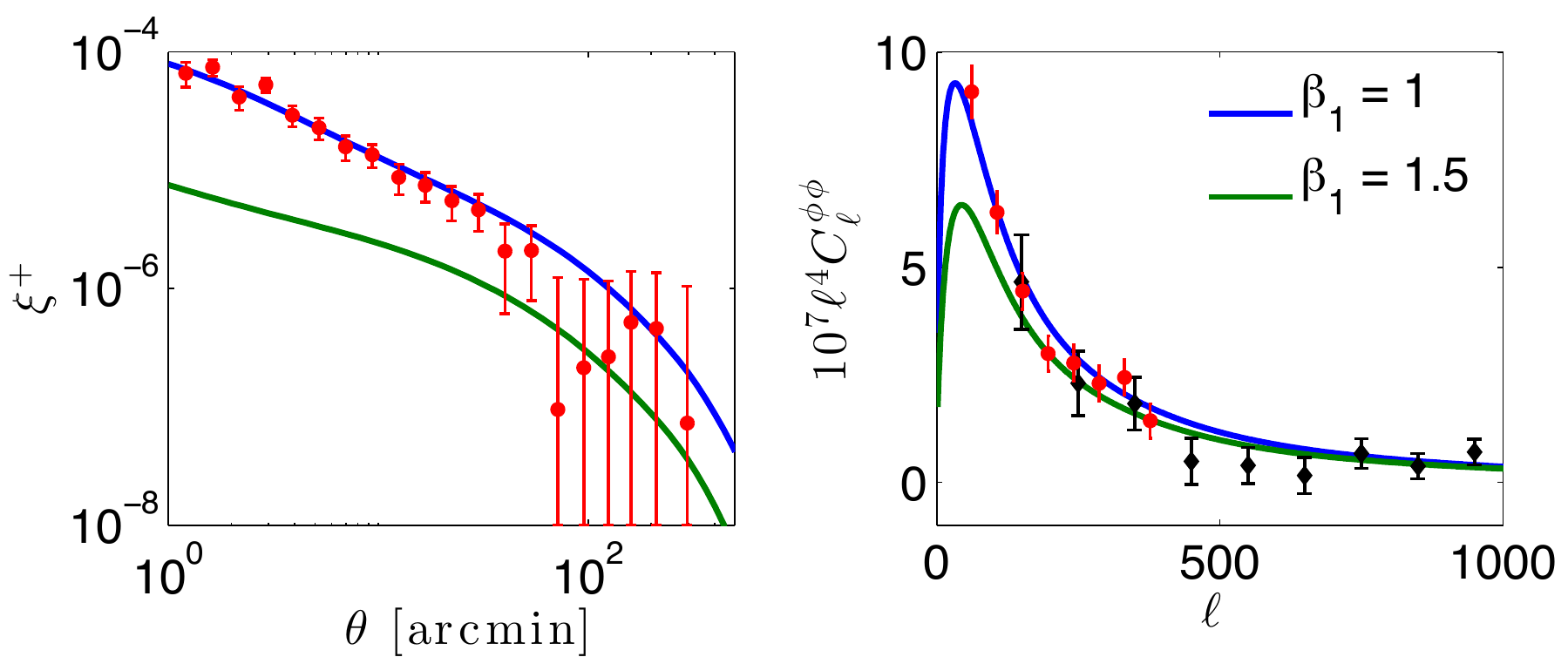}}
      \end{center}
\caption{Sample lensing spectra in the TDI $\ld(g)$ and GSF models with $w=-0.8$. On the left we plot the predictions of the weak lensing correlation $\xi^+$ spectra, as well as the data taken from { {CFHTLenS}}. On the right we plot the lensing spectra of the CMB, $C_{\ell}^{\phi\phi}$, with observational data from \Planck (red/circles) and SPT (black/diamonds).  Here we see that decreasing the sound speed decreases the large-scale clustering of the dark energy.  }\label{fig:COMP_abnew}
\end{figure}

In \fref{fig:COMP_abnew} we plot the predictions for $C_{\ell}^{\phi\phi}$ spectrum and galaxy correlation function $\xi^+$, for the   TDI $\ld(g)$ and  GSF models  (for simplicity we take $\alpha, \beta_1$ to be constants and $\beta_2= 1-\beta_1$, which removes the $\ddot{h}$-term). For galaxy lensing, we use a single redshift bin with $\xi^{\pm} \equiv \xi_{1,1}^{\pm}$ distribution of source galaxies from the Canada-France-Hawaii Telescope Lensing Survey ({CFHTLenS})~\cite{2013MNRAS.430.2200K}. For illustration we have applied corrections from the {\tt Halofit} fitting formulae~\cite{Smith:2002dz,Takahashi:2012em} to the lensing potential $P_{\Psi}$. However, since we do not have a full understanding of  dark energy perturbations on non-linear scales, we do not include those scales where this correction is significant to obtain constraints on the equations of state in the next section.  We compare these predictions to CMB lensing data from   \Planck \cite{2013arXiv1303.5077P} and the South Pole Telescope ({SPT}) \cite{vanEngelen:2012va}, and  also plot the single-bin observations from {CFHTLenS}.  

It is  clear that there will be ranges of parameter space which   yield observational spectra which are compatible with the present data and therefore will not be distinguishable at present.  It is most interesting to note that some of the spectra for CMB lensing are compatible with the data, but are incompatible with the galaxy weak lensing observations: the CFHTLenS galaxy weak lensing data alone could rule out a range of parameter space. 

%
%

\begin{figure}[!t]
      \begin{center}
{{\includegraphics[scale=0.4,angle=0]{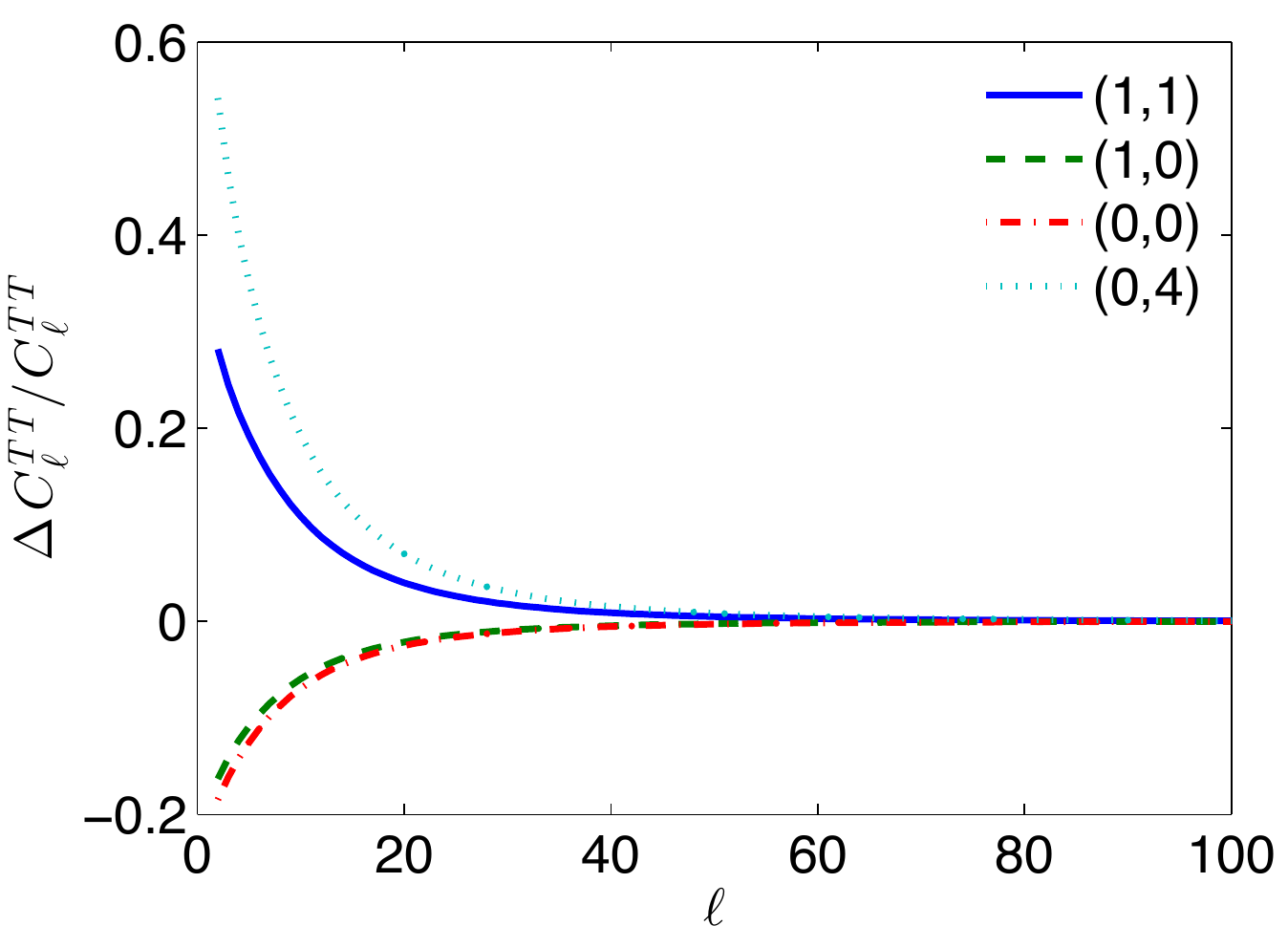}}}
{{\includegraphics[scale=0.4,angle=0]{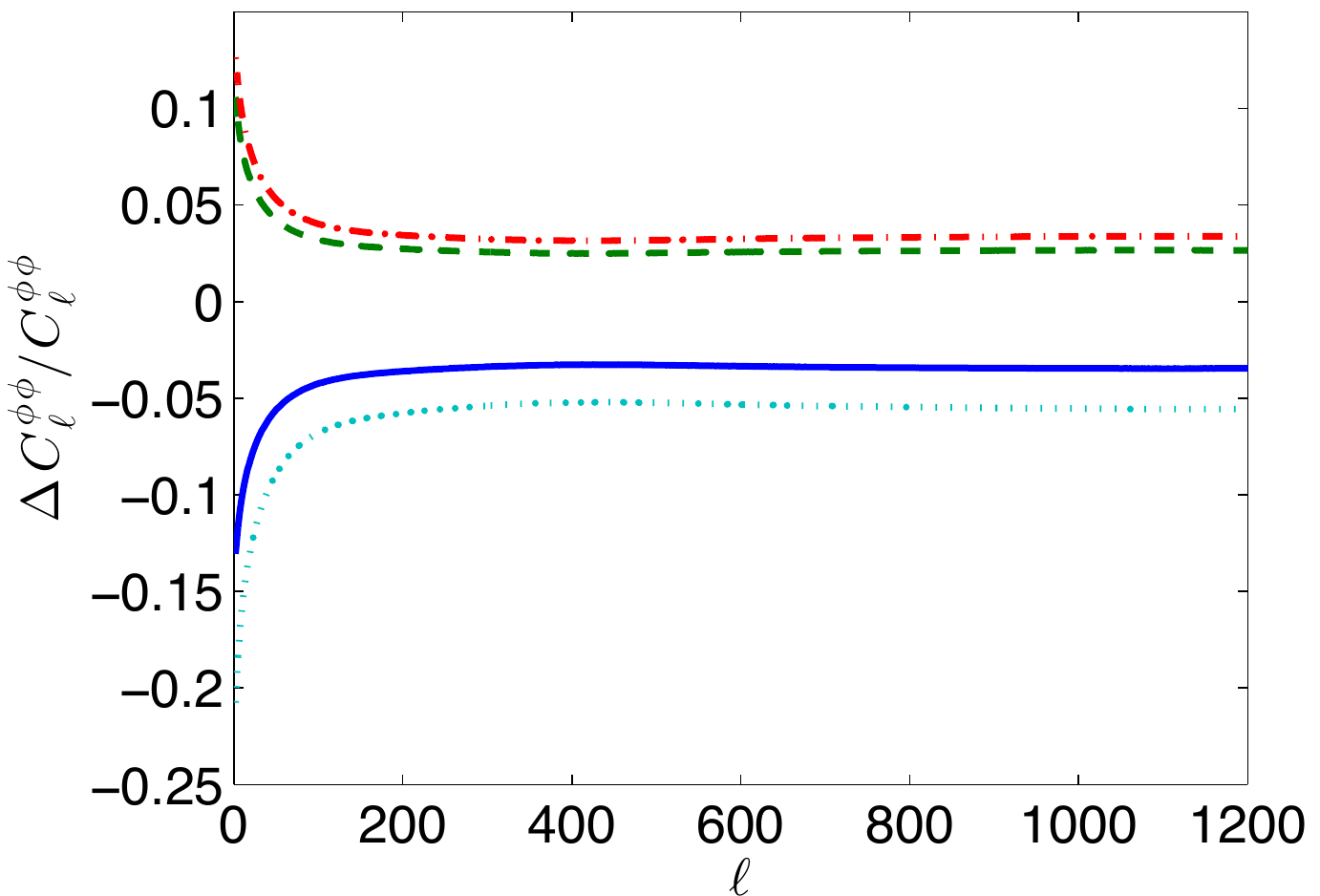}}}\\
\vspace{0.2cm}
{{\includegraphics[scale=0.4,angle=0]{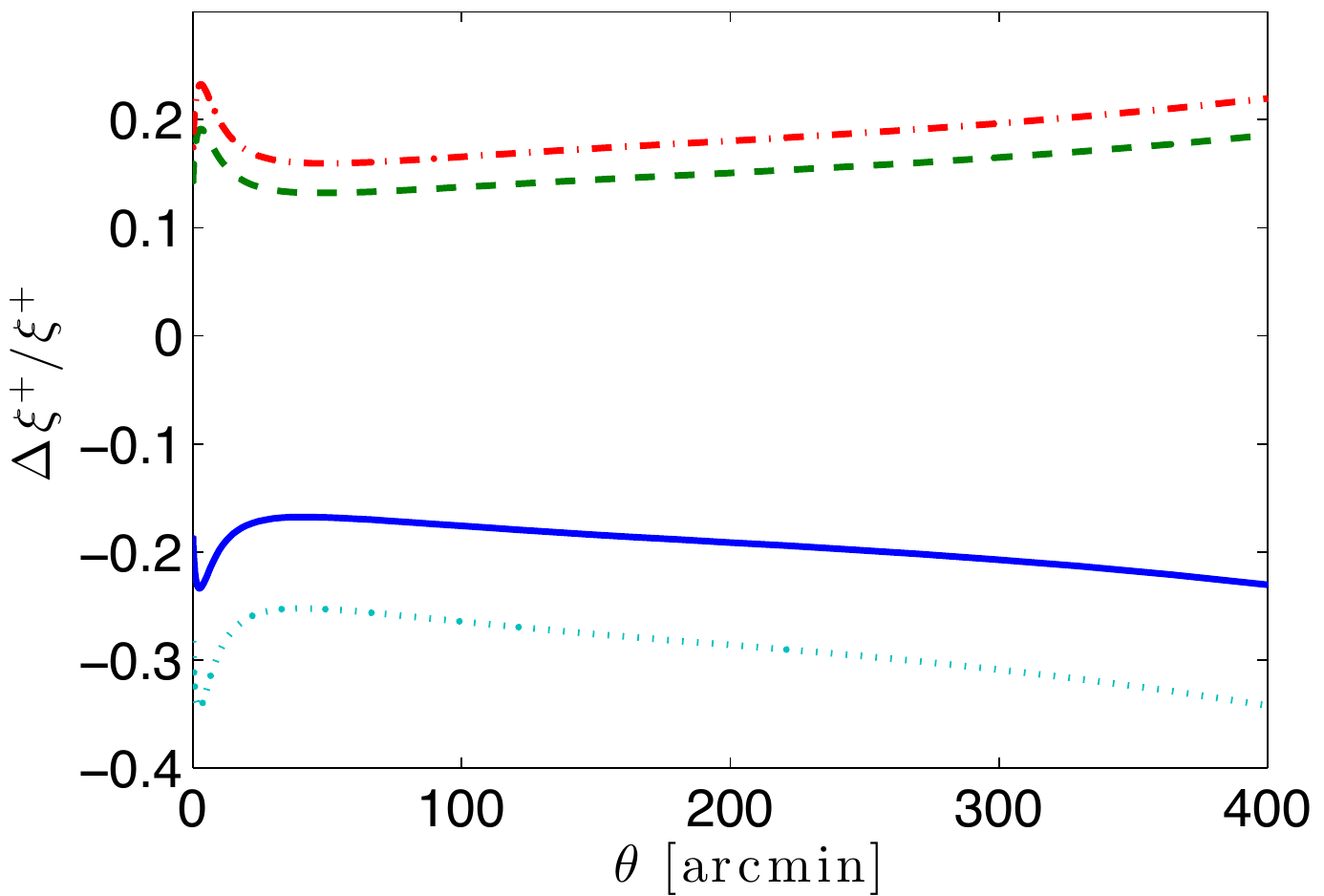}}}
      \end{center}
\caption{The fractional differences in observable spectra for GSF models with  various values of $(\beta_1, \beta_2)$, relative to a  quintessence scenario. $w=-0.8$ and $\alpha=0$ are used throughout. Notice that the fractional shifts in spectra become suppressed for $\ell>50$ in the CMB temperature spectrum $C_{\ell}^{TT}$, remain roughly constant in the lensing of the CMB spectrum, $C_{\ell}^{\phi\phi}$ and grow as a function of $\theta$ for the galaxy weak  lensing correlation function $\xi^+$.   }\label{fig:relative_genab1b2}
\end{figure}

In \fref{fig:relative_genab1b2} we plot the fractional differences in observational spectra for GSF models with a range of values of $\alpha, \beta_1$ and $\beta_2$, relative to a fiducial quintessence scenario with $\alpha=1$, $\beta_1=1$ and $\beta_2=0$. For all of the models shown, the differences in the $C_{\ell}^{TT}$ spectra are negligible for angular scales $\ell >50$, that is, in the noise dominated regime. However,  the differences for the CMB lensing $C_{\ell}^{\phi\phi}$ spectra remain roughly constant for all scales and, markedly, differences in the $\xi^+$ spectra increase with angular size. It is clear that models with an increased power in the  spectrum of  galaxy weak  lensing, $\xi^+$, also have an increased power in the  spectrum of CMB lensing, $C_{\ell}^{\phi\phi}$.  This plot nicely illustrates how useful lensing data is, compared to the CMB temperature data where the differences between the spectra are only significant in the cosmic variance limited regime. The fundamental reason for this is that the lensing data is very sensitive to the evolution of perturbations of the dark sector theory.

\begin{figure}[!t]
      \begin{center}
{\subfigure[\, GSF model with $(\alpha, \beta_1, \beta_2) =(0,1,0)$]{\includegraphics[scale=0.6,angle=0]{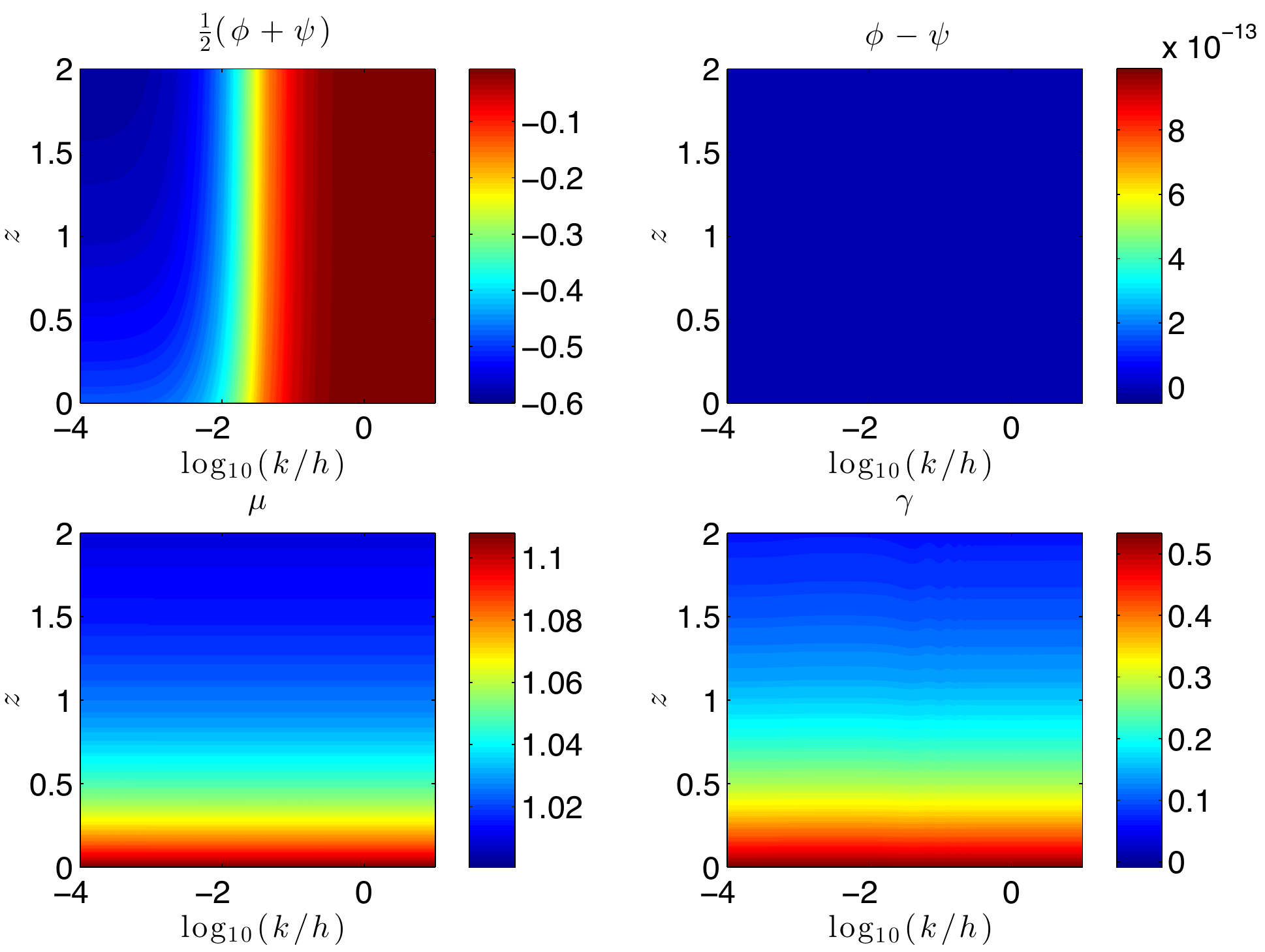}}}
{\subfigure[\, TDI $\ld(g)$ model with $\qsubrm{c}{s}^2 = 10^{-3}$]{\includegraphics[scale=0.6,angle=0]{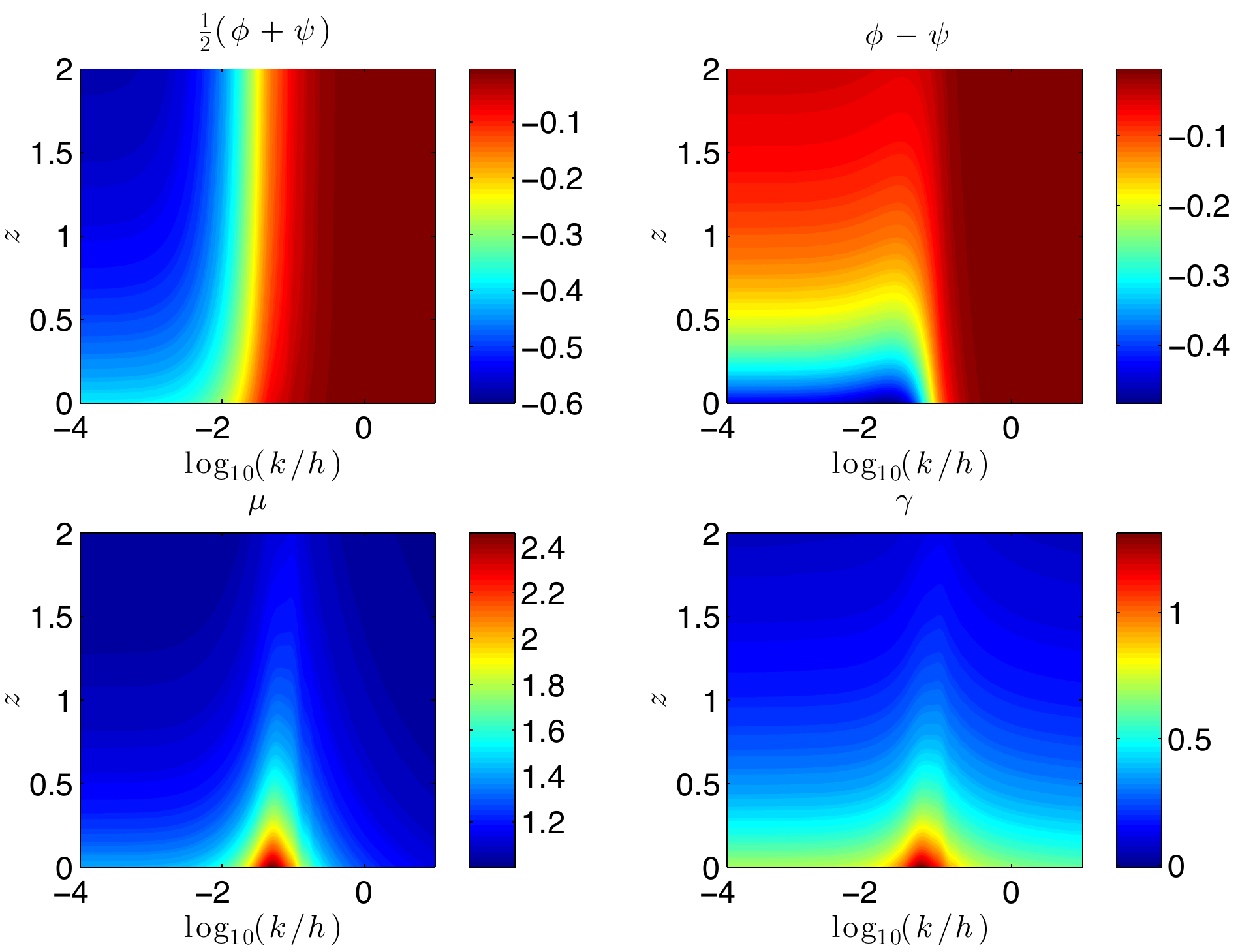}}}
      \end{center}
\caption{Common observational ``diagnostics'', as functions of scale $k/h$ and redshift $z$, for two different models with $w=-0.8$. On the top row we plot the  lensing potential $\psi + \phi$ and stress $\phi-\psi$, and on the bottom row we plot the effective gravitational coupling $\mu$, and growth function $\gamma$.  }\label{fig:obssurfaces}
\end{figure}

It is important to have a handle on which probes are sensitive to which types of theories, and so it is useful to  evaluate other   probes of dark sector theories which are rather prolific in the literature.  We use our modified version of \CAMB to compute the sum and difference of the gravitational potentials, $\phi$ and $\psi$, the effective gravitational coupling $\mu$, defined via
\bea
k^2\psi = -4\pi G\mu a^2\left( \qsubrm{\rho}{k}\qsubrm{\Delta}{k}  + 3 (\qsubrm{\rho}{k}+\qsubrm{P}{k})\qsubrm{\sigma}{k}\right),
\eea
in which $\Delta$ is the comoving density perturbation and a subscript ``k'' denotes a known non-dark-sector component, and the growth function \cite{Wang:1998gt, Linder:2005in} $\gamma$,
\bea
\gamma \defn \frac{1}{\ln\qsubrm{\Omega}{m}}\ln \bigg( \frac{\dd\ln\qsubrm{\delta}{m}}{\dd\ln a}\bigg).
\eea
In \fref{fig:obssurfaces} we plot these quantities as a function of scale and redshift, for (a) GSF models and (b) TDI $\ld(g)$ models. In the GSF models, the growth function $\gamma$ and effective gravitational coupling $\mu$ are close to being scale independent. In contrast, both $\mu$ and  $\gamma$ have clear scale dependence for the  TDI $\ld(g)$ models. This serves to show that a scale-independent parameterization of the gravitational coupling $\mu$ and growth function $\gamma$ would have completely missed the potential observable effects of (in this instance) TDI $\ld(g)$ models, which is a large class of models to be insensitive to.

\begin{figure}[!t]
      \begin{center}
\subfigure[\, $\qsubrm{c}{s}^2 =1$]{\includegraphics[scale=0.6,angle=0]{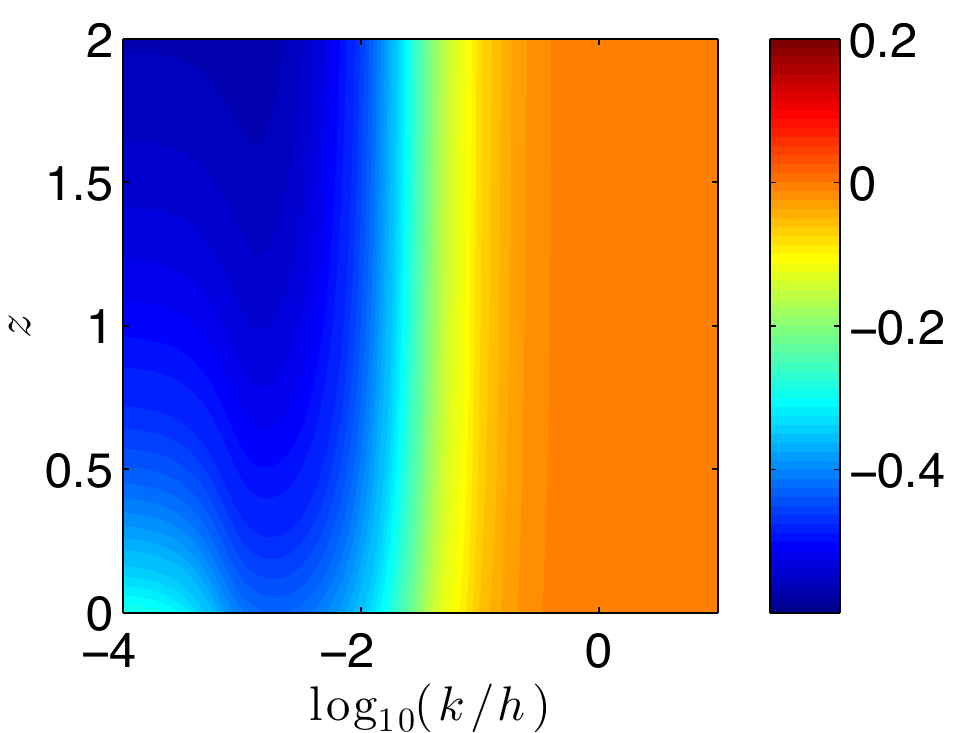}}
\subfigure[\, $\qsubrm{c}{s}^2 =10^{-3}$]{\includegraphics[scale=0.6,angle=0]{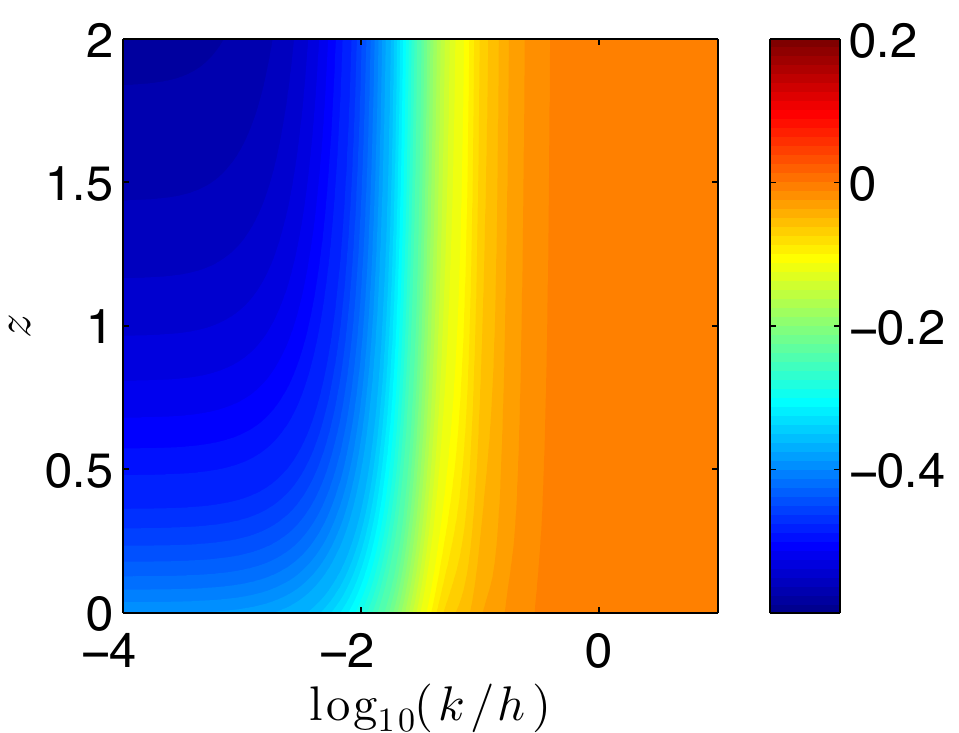}}
\subfigure[\, $\qsubrm{c}{s}^2 =10^{-5}$]{\includegraphics[scale=0.6,angle=0]{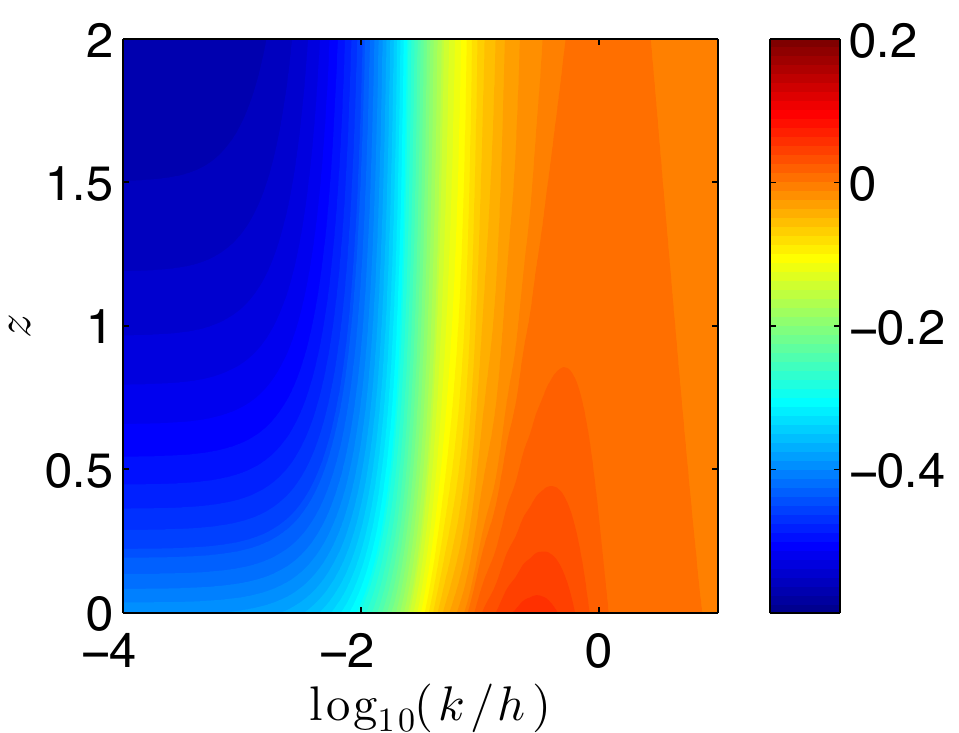}}
\subfigure[\, $\qsubrm{c}{s}^2 =10^{-7}$]{\includegraphics[scale=0.6,angle=0]{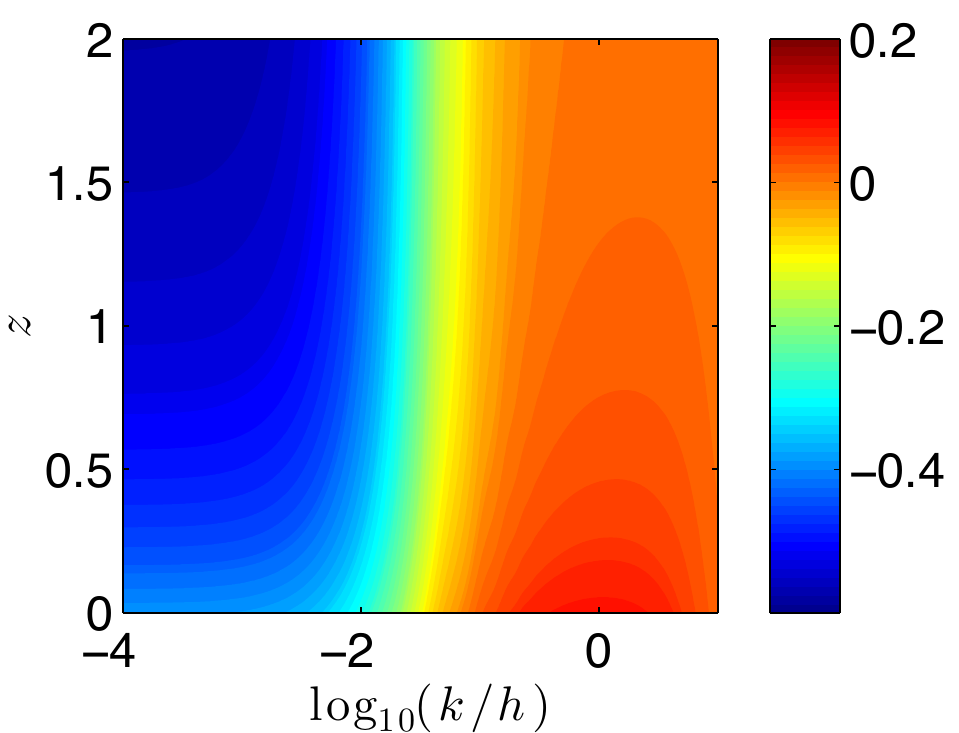}}
      \end{center}
\caption{Lensing potential $\half(\phi+\psi)$ as a function of scale $k/h$ and redshift $z$,  in the $\ld(g)$ TDI models for various values of the sound speed $\qsubrm{c}{s}^2$ (all have $w=-0.8$). The enhacement in lensing at large scales in (a) can be picked up by observations of CMB lensing; this enhancement vanishes in (b). A new enhancement appears at small scales in (c), and has grown in (d);  whilst CMB lensing is insensitive to these scales, galaxy weak lensing becomes sensitive.  }\label{fig:TDI-lensingbehavior}
\end{figure}

In \fref{fig:TDI-lensingbehavior} we plot   the lensing potential in the TDI $\ld(g)$ models, for various values of the sound speed, $\qsubrm{c}{s}^2$. These plots illustrate that the lensing potential has late-time enhancements at different scales depending on the value of $\qsubrm{c}{s}^2$. For large values of $\qsubrm{c}{s}^2$ the lensing potential is enhanced on large scales, and for small values of $\qsubrm{c}{s}^2$ it is enhanced on small scales. This becomes interesting when one realises that CMB lensing and galaxy weak lensing probe the lensing potential on large and small scales, respectively. Hence, data from these two types of experiments can probe different regimes. In the next section we will show our parameter constraints from using current CMB lensing and weak galaxy lensing which corroborates this point. We have performed a similar analysis when varying the $\alpha$ parameter in the GSF models: for $0 \leq \alpha  \leq 1$ we find no qualitative difference in the lensing potential from that we presented in \fref{fig:obssurfaces}(a) -- there certainly aren't any features which pop up on small and large scales as there are in the TDI $\ld(g)$ models. The qualitative difference of the behavior of the lensing potential in the TDI $\ld(g)$ and GSF models appears to be a consequence of the theoretical understanding developed in Section \ref{sec:density}.

\section{Present data constraints}
\label{section:presconstr}
In order to place constraints on the dark sector equations of state, we use  observations of temperature anisotropies in the Cosmic Microwave Background (CMB) from the \Planck satellite~\cite{2013arXiv1303.5062P} and polarisation measurements from the Wilkinson Microwave Anisotropy Probe ({WMAP})~\cite{2012arXiv1212.5225B}. Since the acoustic scale of the CMB is degenerate with the background equation of state $w$ and the Hubble parameter $H_0=100\,h\,{\rm km}\,{\rm sec}^{-1}\,{\rm Mpc}^{-1}$, we use observations of Baryonic Acoustic Oscillations (BAO's) from 6dF~\cite{2011MNRAS.416.3017B}, SDSS (corrected for galaxy peculiar velocities)~\cite{2012MNRAS.427.2132P} and BOSS~\cite{2013arXiv1303.4666A} to constrain the expansion history and exclude models with $w$ far from $-1$ as done in the Planck analysis~\cite{Ade:2013zuv}.

The dark sector perturbations are constrained by the large angle Integrated Sach Wolfe (ISW) component of the primary CMB, and gravitational lensing along the line of sight.  We use CMB lensing as detected by {\it Planck}~\cite{2013arXiv1303.5077P}, {SPT}~\cite{vanEngelen:2012va}, and also from weak galaxy lensing detected by {CFHTLenS}~\cite{2012MNRAS.427..146H,2013MNRAS.430.2200K} to constrain the latter.

The base cosmology we consider is a 7 parameter ${\bf p}=\{\Omega_{\rm b}h^2, \Omega_{\rm c}h^2, \theta_{\rm MC}, A_{\rm S}, n_{\rm S}, \tau, w\}$ model. Here $\Omega_{\rm b}$ and $\Omega_{\rm c}$ are the baryonic and cold dark matter densities relative to the critical density, $\theta_{\rm MC}$ is the acoustic scale of the CMB, $A_{\rm S}$ and $n_{\rm S}$ are the amplitude and spectral index of initial fluctuations respectively, and $\tau$ the optical depth to reionization.  We consider both the TDI $\mathcal{L} (g)$ model, which is has one additional parameter $c_{\rm s}^2$, and the GSF model with additional parameters  $\{ \alpha, \beta_1, \beta_2\}$.

We make use of the {\it Planck} likelihood~\cite{2013arXiv1303.5075P} and follow the same procedure as the {\it Planck}  cosmology analysis~\cite{Ade:2013zuv}, that includes a number of nuisance parameters describing the contamination from our own galaxy, extragalactic sources and the SZ effect.  We consider three data combinations, 
\begin{itemize}
\item[I] {\it Planck} CMB+WP+BAO+CMB lensing, 
\item[II] {\it Planck} CMB+WP+BAO+{CFHTLenS}, and 
\item[III] {\it Planck}+WP+BAO+CFHTLenS+CMB lensing.  
\end{itemize}

For {CFHTLenS} we use the shear correlation functions and covariance matrix as described in~\cite{2013MNRAS.430.2200K}. However, we make a conservative choice of angular scales to include in the analysis. Since we do not have a full understanding of how the dark energy models considered here may cluster on non-linear scales, we do not include those scales which contribute significantly to the shear correlation  in the likelihood. In practice, we exclude the entire $\xi^{-}$  data, and only include $\xi^{+}$ for $\theta > 12$ arcminutes.  This exclusion regime was obtained by switching non-linear corrections on and off, to find the range of $\theta$ for which the $\xi^+$ spectrum is unaffected. This eliminates the need to apply any non-linear corrections, such as {\tt Halofit}.  For {SPT} lensing data we follow the same procedure as in~\cite{vanEngelen:2012va}, rescaling the diagonals elements of the covariance matrix according to sample variance, and adding an additional calibration-induced uncertainty to the covariance. The non-linear correction to the CMB lensing is small enough to safely ignore it. 

Parameter constraints were obtained from Markov Chain Monte Carlo (MCMC) chains produced using the {\tt COSMOMC} code~\cite{Lewis:2002ah}. Prior ranges on parameters were chosen to sufficiently contain the relevant posterior distributions. For the dark energy equations of state, these are given in Table~\ref{tab:priors}. We note that, formally, the equations of state that we have used are not defined for $w<-1$, but solving the equations of motion appears to give stable solutions and hence we include them as a ``continuation'' of physical models. Convergence of the MCMC chains was ensured by terminating the runs when the Gelman and Rubin {\tt R-1} value, characterising the variance between chains, was $<0.05.$

\begin{table}[!t]
\caption{Prior ranges on the dark energy equations of state.}
\label{tab:priors}
\begin{center}
\begin{tabular}{|c|c|}  \hline 
Parameter & Prior range \\ \hline 
$w$ & [-1.5, -0.5]\\ \hline \hline
$\log_{10} c_{\rm s}^2$ & [-4, 0] \\ \hline \hline
$\log_{10} \alpha$ & [-5, 0] \\ \hline 
$\beta_1$ & [0, 1.9]\\ \hline 
$\beta_2$ & [0, 15] \\ \hline  \hline 
\end{tabular}
\end{center}
\label{default}
\end{table}%

\begin{figure}[!t]
      \begin{center}
{{\includegraphics[scale=0.8,angle=0]{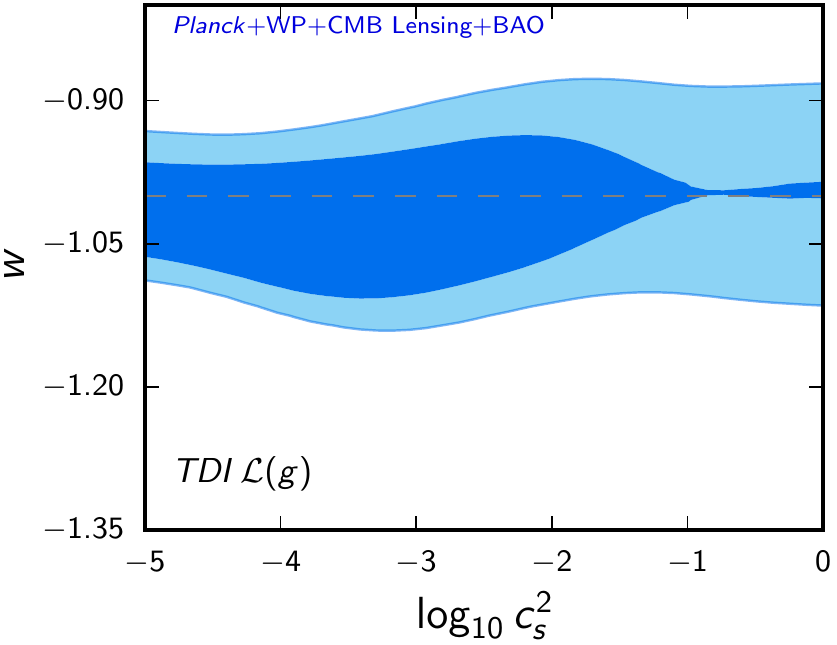}}}
{{\includegraphics[scale=0.8,angle=0]{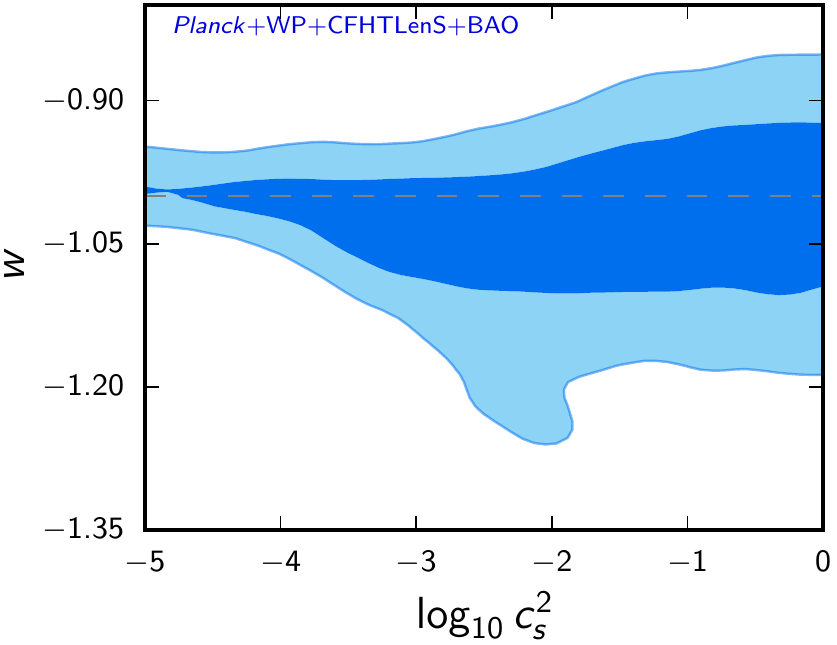}}}\\
\vspace{0.2cm}
{{\includegraphics[scale=0.8,angle=0]{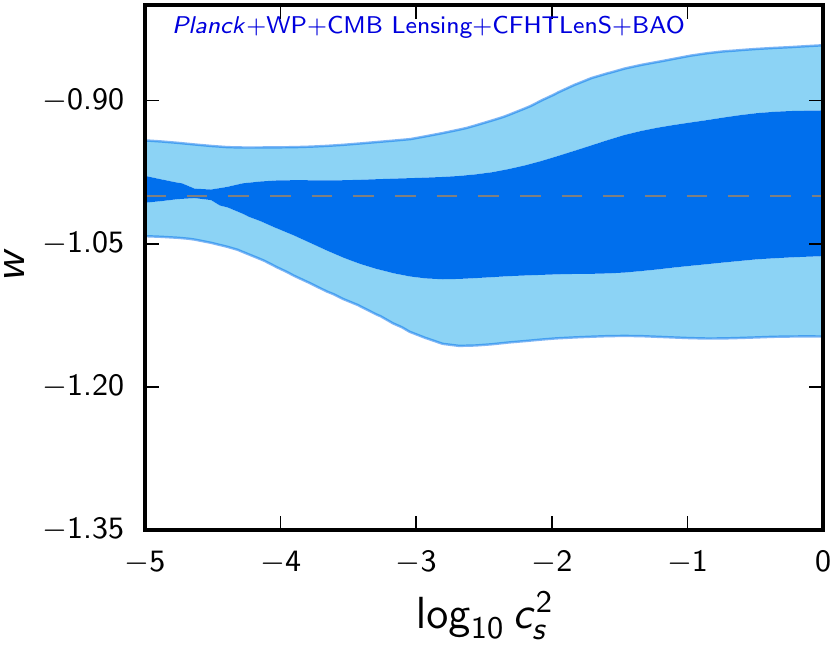}}}
      \end{center}
\caption{\, Constraints on the (background) equation of state parameter $w$ and the squared sound speed $c_{\rm s}^2 $ for the TDI $\mathcal{L}(g)$ model for three different data combinations. Data used in the analysis are \Planck CMB temperature and WMAP polarization, BAO, together with either  CMB lensing from \Planck and {SPT} (top left panel, I),  galaxy lensing from {{CFHTLenS}}  (top right panel, II), and a combination of CMB lensing from \Planck and galaxy lensing from {{CFHTLenS}}  (bottom panel, III).  {{CFHTLenS}}  data is truncated to the linear regime, as discussed in the text. }\label{fig:ede_planck_cfhtlens}
\end{figure}

\begin{figure}[!t]
      \begin{center}
{{\includegraphics[scale=0.8,angle=0]{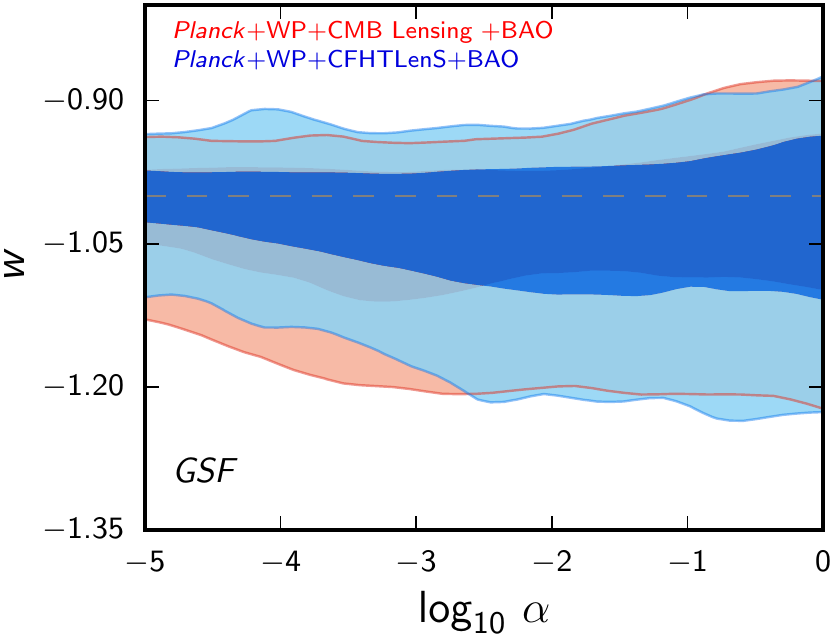}} 
{\includegraphics[scale=0.8,angle=0]{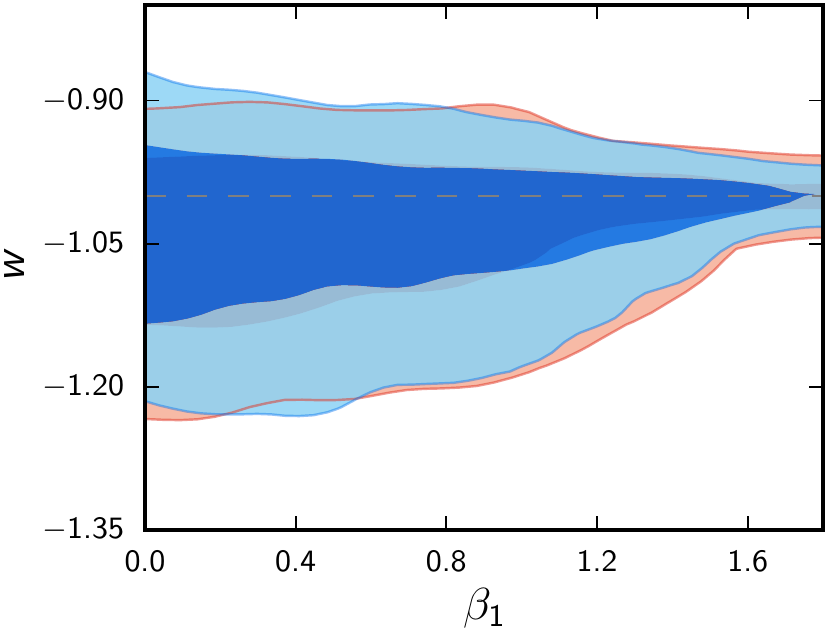}}\\
\vspace{0.2cm}
 {\includegraphics[scale=0.8,angle=0]{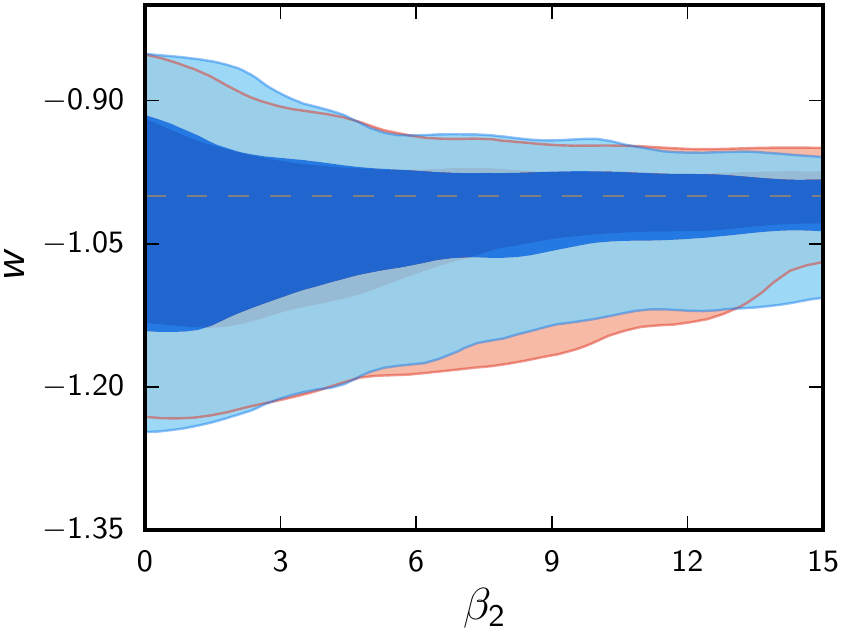}}}
      \end{center}
\caption{\, Constraints on $w$ and $\alpha, \beta_1, \beta_2$ for the generalised scalar field model. Labelling is the same as in Fig.~\ref{fig:ede_planck_cfhtlens}. }\label{fig:phi_planck_cfhtlens}
\end{figure}

2D posterior  likelihoods are shown in Figs.~\ref{fig:ede_planck_cfhtlens} and~\ref{fig:phi_planck_cfhtlens} for the two dark energy parameterizations we have considered. Constraints on the background equation of state $w$ are consistent with previous studies combining CMB + BAO data that use a different prescription for dealing with the perturbations~\cite{PhysRevD.76.104043, PhysRevD.77.103524, Fang:2008sn}. However, one of the encouraging points of this paper  is that {\em current} observations can also constrain the parameters of the perturbative  equations of state.  In the TDI $\mathcal{L} (g)$ model, shown in Fig.~\ref{fig:ede_planck_cfhtlens}, a low sound speed results in clustering of the dark energy fluid on scales probed by observations. However, since in the limit of $w \rightarrow -1$  the fluid behaves like $\Lambda$, which is smooth with no perturbations, one observes a narrow valley around $w=-1$ in which $c_{\rm s}^2$ cannot be constrained. 

Similar behaviour is also observed in the generalised scalar field model, shown in Fig.~\ref{fig:phi_planck_cfhtlens}. In this case $\alpha$, which acts as the sound speed in quintessence models, does not result as strong clustering even when $\alpha \rightarrow 0$ as in the case of TDI $\mathcal{L} (g)$, implying that it cannot be constrained by current observations. However, varying $\beta_1$ and $\beta_2$ can lead to observable effects as shown in Fig.~\ref{fig:COMP_abnew} and hence be constrained, as long as $w$ is  sufficiently far from $w=-1$.

\begin{figure}[!t]
      \begin{center}
{{\includegraphics[scale=0.8,angle=0]{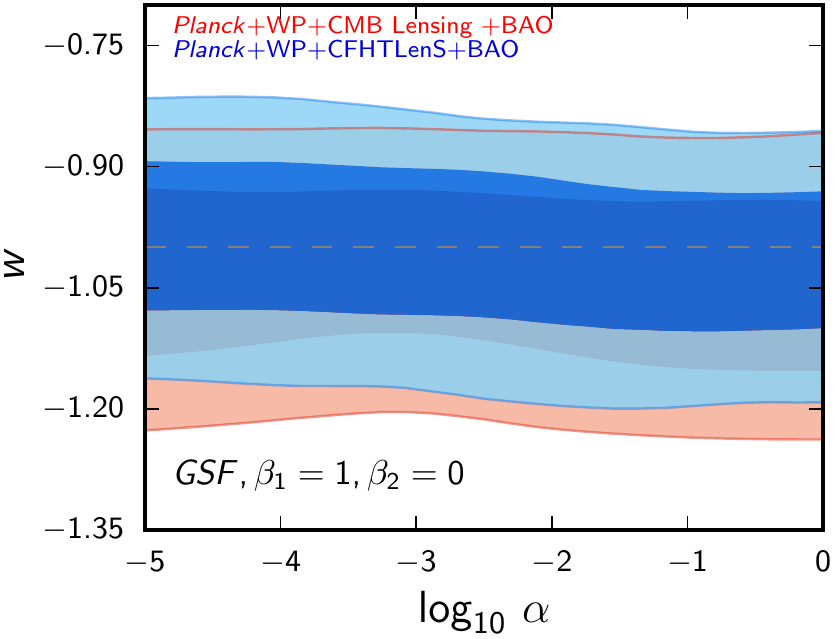}}}
      \end{center}
\caption{\, Constraints on $w$ and $\alpha$ for the generalised scalar field model with $\beta_1=1, \beta_2=0$: this correspond to models of $k$-essence type. Note that the constraint on $w$ appears to be almost independent of $\alpha$.}\label{fig:phi_b12_planck_cfhtlens}
\end{figure}

There are two interesting models which are included within our GSF class of models. The first are all models whose Lagrangian is $\ld = \ld(\phi, \kin)$: as remarked above, these models have $\beta_1 =1, \beta_2=0$, and $\alpha$ (which in general is a function of time) is the only free function which controls the evolution of perturbations. In \fref{fig:phi_b12_planck_cfhtlens} we present constraints on $w$ and $\alpha$ for these models (both are assumed to be constant). It is apparent that constraints on $\alpha$ are almost independent of $\alpha$. One should compare these constraints with the top panel of \fref{fig:phi_planck_cfhtlens}, which are for the full GSF model parameters: those show some non-trivial constraints on low values of $\alpha$. The differences are non-trivial impact of the constraints on $\beta_1$ and $\beta_2$.

\begin{figure}[!t]
      \begin{center}
{{\includegraphics[scale=0.4,angle=0]{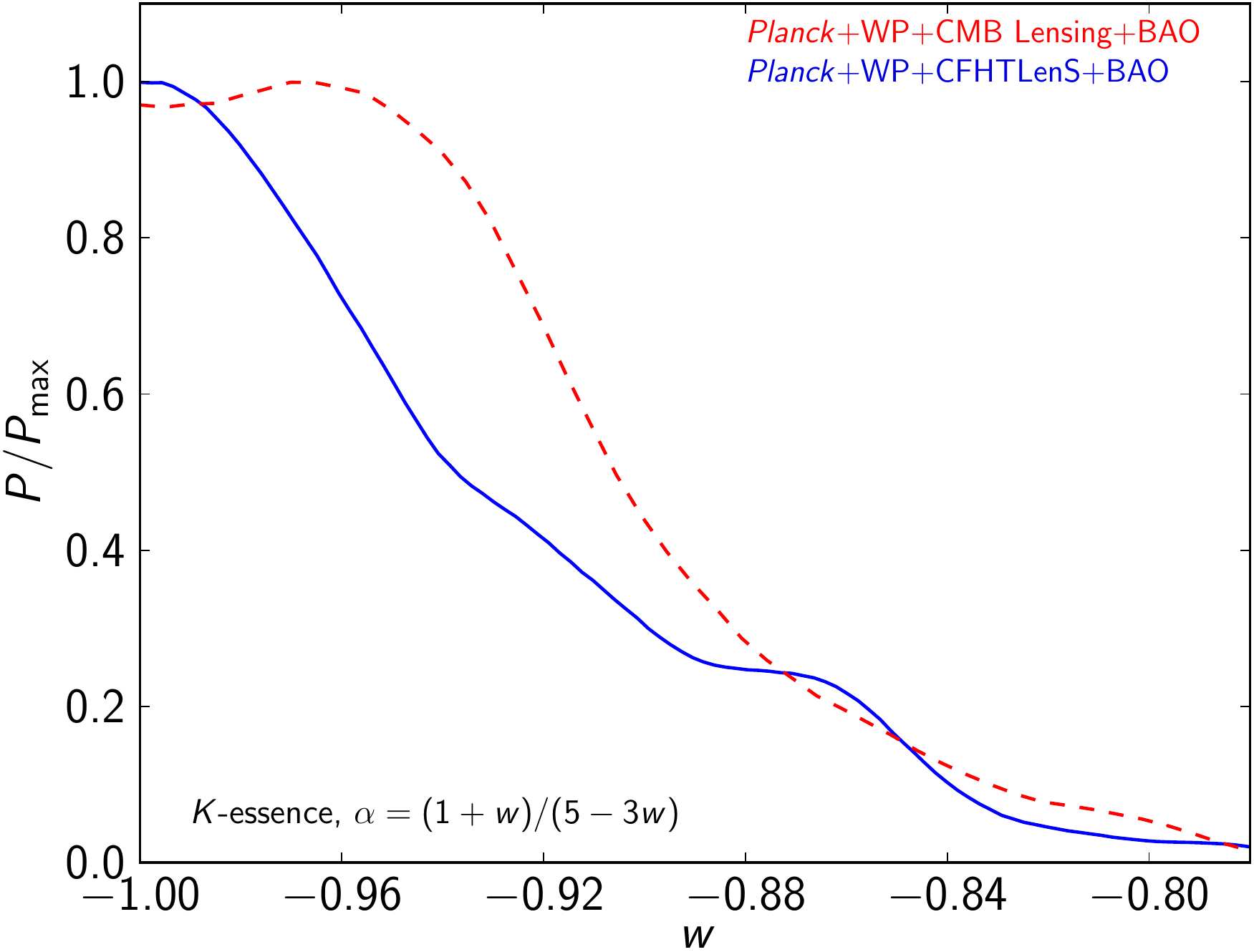}}}
      \end{center}
\caption{Constraints on the equation of state parameter, $w$, for k-essence models with the condition that $\alpha = (1+w)/(5-3w)$, that is, for the model presented in (\ref{kess1}) and (\ref{kess2}). Note that it is not possible for $w$ to be $<-1$ in this models. This is a 1D section through the 2D likelihood presented in Fig.~\ref{fig:phi_b12_planck_cfhtlens}} \label{fig:ke_planck_cfhtlens-1d}
\end{figure}

The second interesting subset of models are those with $\alpha = (1+w)/(5-3w), \beta_1=1, \beta_2=0$ which were explained at the end of section \ref{sec:gsf}; constraints on $w$ are presented in \fref{fig:ke_planck_cfhtlens-1d}.

\subsection{Constraints with prejudices and priors on $w$}
The parameters in the equations of state for perturbations become difficult to constrain when $w\approx -1$. However, it is apparent from the constraints shown in Figs.~\ref{fig:ede_planck_cfhtlens} and \ref{fig:phi_planck_cfhtlens} that there are some preferred values of the parameters when $w\neq -1$. 
It is conceivable that an  observation of background cosmology alone will constrain $w$. If the central value of this constraint lies away from $w = -1$ then observations of the perturbed Universe can be used to place  more   stringent bounds on the parameters in the equations of state for perturbations than if $w$ is ``totally free'', as it has been in the previous section. 

First, suppose that the data preferred some value of $w$; for concreteness we will pick $w = - 0.95 \pm 0.01$. Then the constraints on the GSF parameters are: $\beta_1 < 1.3$ for all data combinations, and $\beta_2 < 13$ and $\beta_2 < 11$ for the $Planck$+WP+BAO+CFHTLenS and $Planck$+WP+BAO+CMB lensing data combinations respectively. In the TDI $\ld(g)$ case only the former data combination yields a useful constraint, and we find $\log_{10}\qsubrm{c}{s}^2 < -4$ is ruled out.

\begin{figure}[!t]
      \begin{center}
{{\includegraphics[scale=0.4,angle=0]{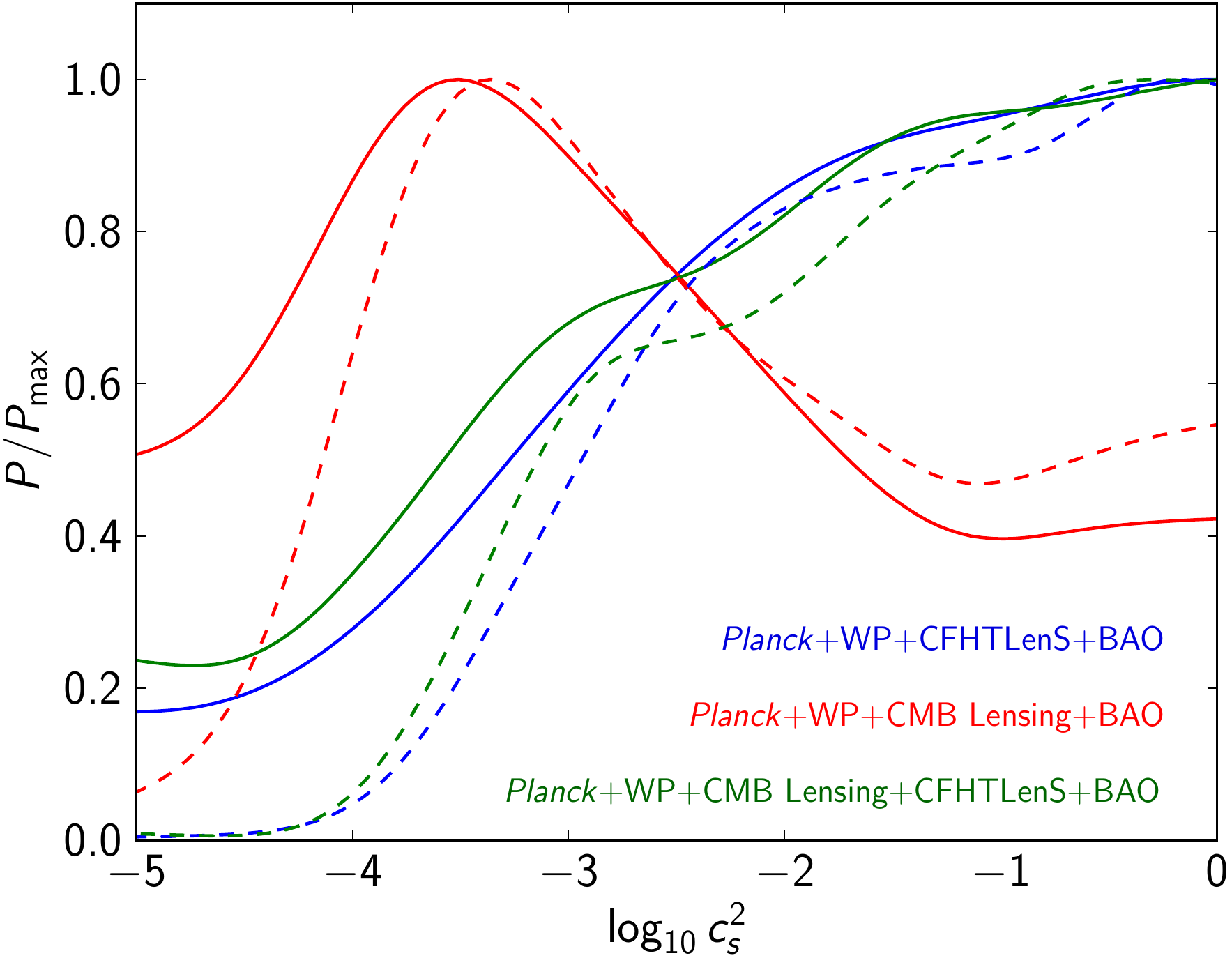}}}
      \end{center}
\caption{Marginalized 1D likelihoods for the sound speed parameter, $\qsubrm{c}{s}^2$, for the TDI $\ld(g)$ model. The solid lines have no prior on $w$ (other than that in Table \ref{tab:priors}), and the dashed lines  have the prior $|1+w| > 0.05$.  }\label{fig:ede_planck_cfhtlens-1d}
\end{figure}

\begin{figure}[!t]
      \begin{center}
{{\includegraphics[scale=0.4,angle=0]{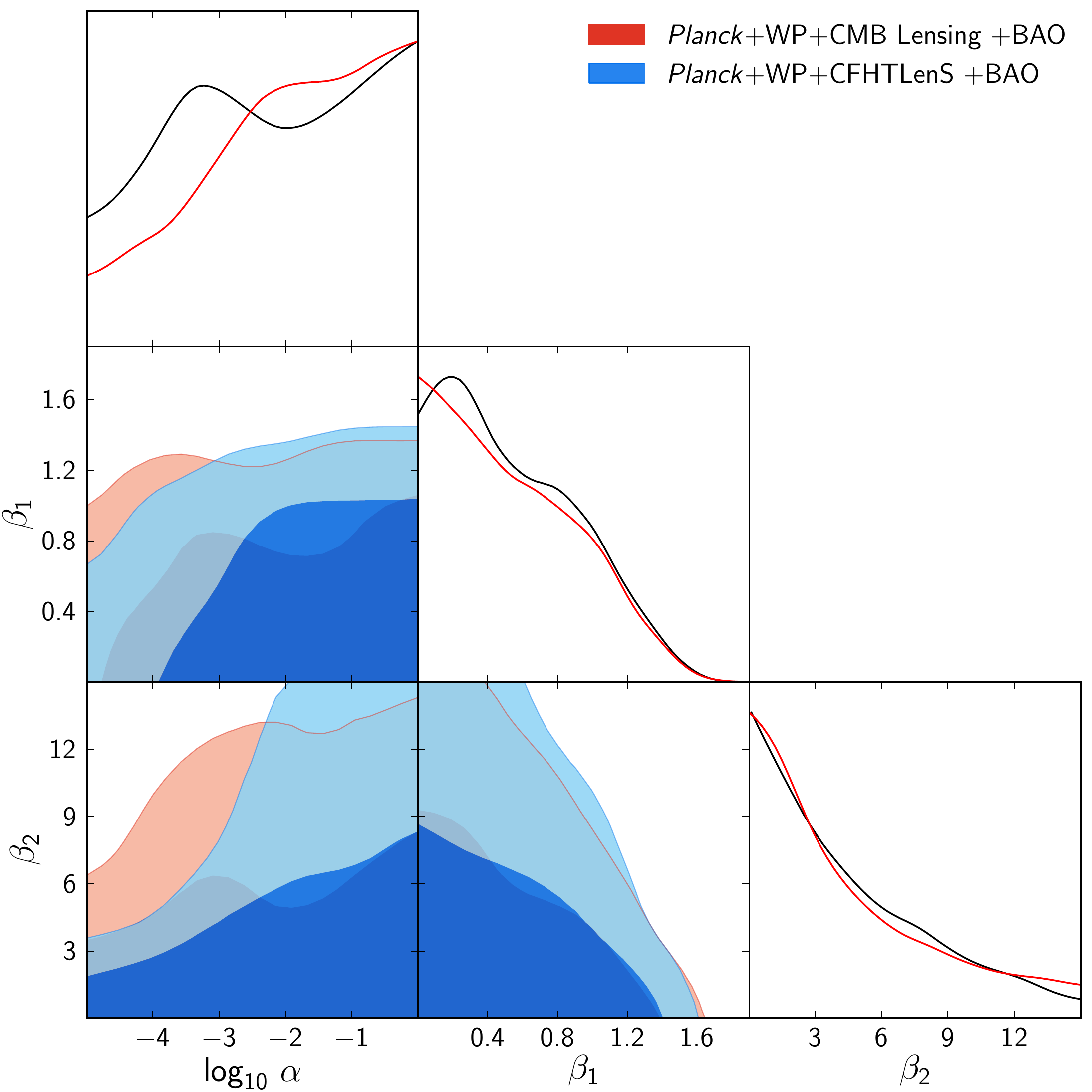}} }
      \end{center}
\caption{\, Constraints on $\alpha, \beta_1, \beta_2$ for the GSF model,  using a prior of $|1+w|>0.05$. Constraints are qualitatively similar without using this prior.}\label{fig:phi-wprior}
\end{figure}

Secondly, by applying the prior $|1+w|> 0.05$ we can remove the region of parameter space in which there is little clustering. In Fig.~\ref{fig:ede_planck_cfhtlens-1d} we show the constraints on $\qsubrm{c}{s}^2$ in the TDI $\ld(g)$ models and in Fig.~\ref{fig:phi-wprior} the constraints on $\alpha, \beta_1, \beta_2$ in the GSF models. After inspecting \fref{fig:phi-wprior}, our  result for the GSF model is that the following ranges of parameter space are ruled out:
\bse
\bea
\beta_1 > 1.47 - 0.03(\log_{10}\alpha)^2,
\eea
\bea
\beta_2 > 33.5 + 11.5 \log_{10}\alpha + 1.1 (\log_{10}\alpha)^2,
\eea
\bea
\beta_2 > 15(1.6 - \beta_1).
\eea
\ese
In future work we will  compute $\{\alpha, \beta_1, \beta_2\}$ for specific models.

\section{Future constraints}
\label{section:futureconstr}
The aim in this section is to determine how well  upcoming experiments, such as {\it Euclid}  \cite{Amendola:2012ys},  {PRISM} \cite{Andre:2013afa}, and {CoRE} \cite{Bouchet:2011ck}, might be able to distinguish different values of the parameters in the equations of state for perturbations. This will provide an opportunity to compare the effectiveness of CMB experiments against galaxy weak lensing experiments, and how improvements in the understanding of non-linear physics could strengthen constraints.

In order to do this, we will compare the observational spectra from two different models: one of these will be our fiducial model, which we take to be quintessence. The other model will be  either a TDI $\ld(g)$ model with a given value of $\qsubrm{c}{s}^2$, or a GSF model with a given combination of the parameter values $\{\alpha, \beta_1, \beta_2\}$. We will  compute the $\chi^2$ between the spectra of  the two models. This will be done for CMB-type experiments, where the data included are the temperature, polarization, and CMB lensing  spectra, and tomographic galaxy weak lensing experiments, where the data included is the convergence power spectra correlated across different redshift bins. The    $\chi^2$ will indicate the potential discriminatory power; below we explain how $\chi^2$ is computed for the CMB and tomographic galaxy weak lensing experiments.


At each multipole $\ell$-mode we compute the $n\times n$ covariance matrix $P_{ij}(\ell)$ for a given type of observation (e.g. CMB or galaxy weak lensing). The covariance matrices for two models, $\widehat{P}_{ij}(\ell)$ and $\overline{P}_{ij}(\ell)$ say, are modified to include experimental noise:
\bea
 {P}_{ij}(\ell)\longrightarrow  {P}_{ij}(\ell) + \mathcal{N}_{ij}(\ell).
\eea
The  properties of the noise,  $\mathcal{N}_{ij}(\ell)$, are experiment and survey dependent, but we assume they are  known and diagonal.  Computing the $\chi^2\defn - 2\ln\ld$ between the two models   yields 
\bea
\label{eq:sec:chisq-howtocompute}
\chi^2 = \sum_{\ell=2}^{\qsubrm{\ell}{max}}\qsubrm{f}{sky}(2\ell+1)\Delta \ell\bigg[ \widehat{P}_{ij}^{-1}\overline{P}^{ij} + \ln\left| \frac{|\widehat{P}|}{| \overline{P}|}\right| - n \bigg],
\eea
where $\qsubrm{f}{sky}$ is the fraction of the sky covered by a given survey, $\Delta \ell$ is the width of the multipole bin,  $ \widehat{P}_{ij}^{-1}\overline{P}^{ij} \equiv \sum_{i,j=1}^n \widehat{P}_{ij}^{-1}\overline{P}^{ij}$, $|P|$ denotes the determinant of  $P$, and $n$ is the dimension of the covariance matrix.   For the weak galaxy lensing experiments, we consider a range of upper-limits $\qsubrm{\ell}{max}$ in (\ref{eq:sec:chisq-howtocompute}) to represent our understanding (or lack-of) of the dark sector on non-linear scales.  These are $\qsubrm{\ell}{max}= 10^2$, for which non-linear effects are negligible, and $\qsubrm{\ell}{max}=10^3$, for which non-linear effects come into play, but we might hope that there exists some understanding (in practice, we use {\tt Halofit}).

For CMB experiments, the covariance matrix is constructed from the temperature $T$, curl-free polarization $E$, and  deflection $d$ correlation functions $ {C}^{XY}_{\ell}$ according to
\bea
\qsuprm{P}{CMB}_{ij}(\ell) = \left( \begin{array}{ccc} C^{TT}_{\ell} & C^{TE}_{\ell} & C^{Td}_{\ell} \\ C^{TE}_{\ell} & C^{EE}_{\ell} & 0 \\ C^{Td}_{\ell} & 0 & C^{dd}_{\ell}\end{array} \right).
\eea
This construction is identical to that given in \cite{Perotto:2006rj}.
For galaxy weak lensing observations, the covariance matrix is computed via (\ref{eq:sec:weak-lensing-ij-conv}), and describes the correlations of the convergence spectra between $N$ redshift bins. 

For the CMB, we assume noise excepted for   {CoRE} \cite{Bouchet:2011ck}, {PRISM} \cite{Andre:2013afa},  \Planck \cite{citeulike:11697427} and a hypothetical cosmic variance limited experiment.  For tomographic galaxy weak lensing, we use the {\it Euclid} RedBook\footnote{Also see the {\it Euclid} website, \url{http://www.euclid-ec.org}.} noise values \cite{Amendola:2007rr}, which are given in Table.~\ref{tab:euclid}. Here the noise term is 
\bea
\mathcal{N}_{ij} = \delta_{ij}\expec{\qsubrm{\gamma}{int}^2}n_j^{-1},
\eea
where $\qsubrm{\gamma}{int}$ is the rms intrinsic shear, $n_i$ is the number of galaxies per steradians in the $\qsuprm{i}{th}$-bin, given by
\bea
 n_i = 3600\, d\, (180/\pi)^2\hat{n}_i\,,
\eea
where $d$ is the number of galaxies per square arcminute, and $\hat{n}_i$ is the fraction of sources in the $\qsuprm{i}{th}$-bin. We assume a radial distribution of sources to a maximum redshift $z_{\rm max}$, given by 
\bea
d(z) = z^2 \exp \left[- (z/z_0)^{1.5} \right]\,,
\eea
constructed in such a way that there are an equal number of galaxies for each of the  $N$ redshift bins. For {\it Euclid}, this means a constant $1.17\times 10^{-10}N$ is added to the diagonal elements of the covariance matrix. 

\begin{table}[!t]
\caption{{\it Euclid} survey parameters.}
\begin{center}
\begin{tabular}{|c|c|} \hline
Parameter & Value \\ \hline
 $\expec{\qsubrm{\gamma}{int}}^{1/2}$ & 0.22 \\ \hline
$d$ & 33 \\ \hline
$N$ & 10 \\ \hline
$z_{\rm max}$ & 2.5 \\ \hline
$z_0$ & 1.0 \\ \hline
\end{tabular}
\end{center}
\label{tab:euclid}
\end{table}%

\begin{figure}[!t]
	\begin{center}
		{\includegraphics[scale=0.8,angle=0]{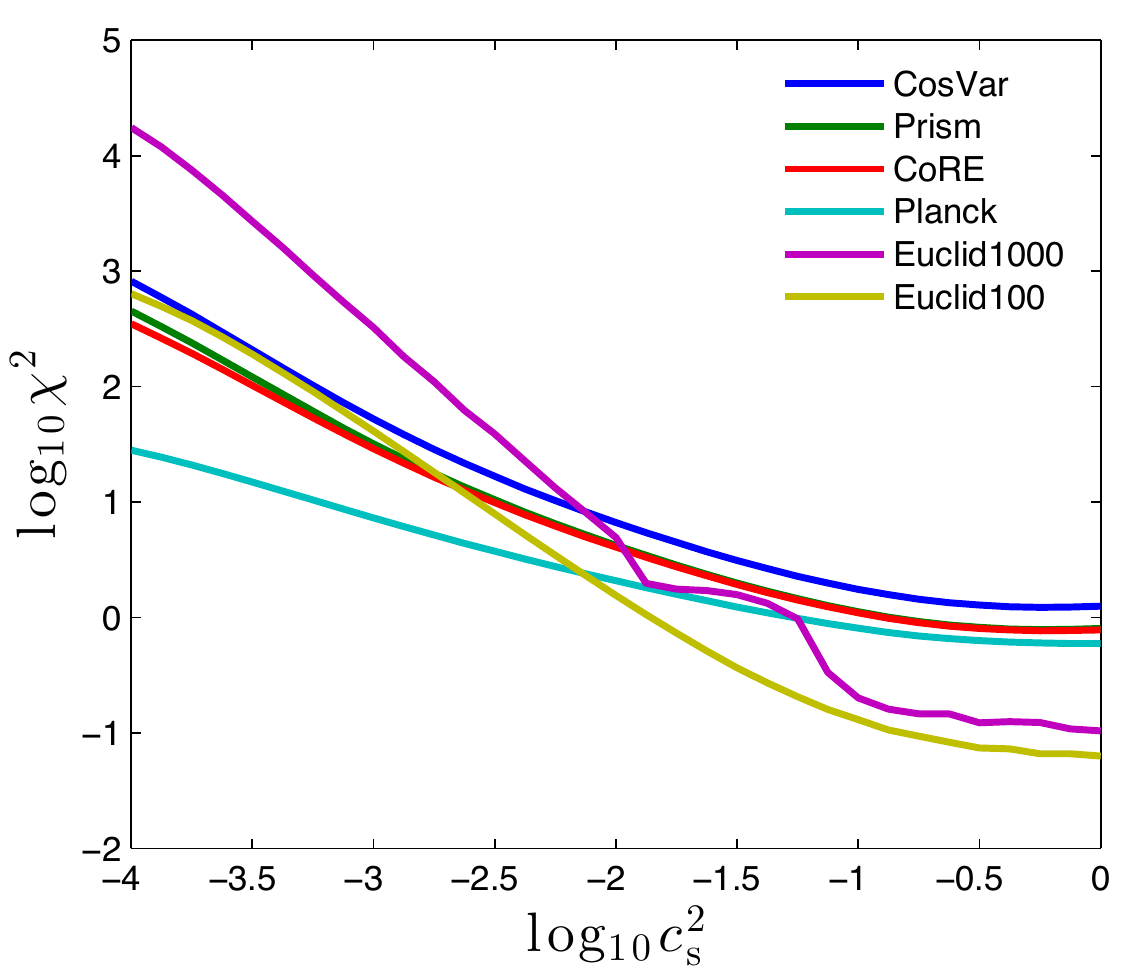}}
      	\end{center}
\caption{Comparing the discriminatory power of CMB and galaxy weak lensing experiments forthe sound speed parameter $\qsubrm{c}{s}^2$ in the TDI $\ld(g)$ models. We plot   $\chi^2$ for CMB experiments with cosmic variance,  {PRISM}, {CoRE}, and \Planck noises, and tomographic galaxy weak lensing measurements using {\it Euclid} survey parameters with $\qsubrm{\ell}{max} = 10^2, 10^3$ respectively.  We have used a fiducial cosmological model with $w=-0.8$.  }\label{fig:chisq_gen-tdilg}
\end{figure}

In \fref{fig:chisq_gen-tdilg} we plot the $\chi^2$ values for the $\ld(g)$ TDI models for  CMB and tomographic weak lensing experiments. It is perhaps interesting to note that the tomographic weak lensing surveys alone will allow us to probe very low values of the sound speed parameter. The level of the constraining power clearly increases as $\qsubrm{\ell}{max}$ is increased.

\begin{figure}[!t]
	\begin{center}
		\subfigure[\, PRISM]{\includegraphics[scale=0.5,angle=0]{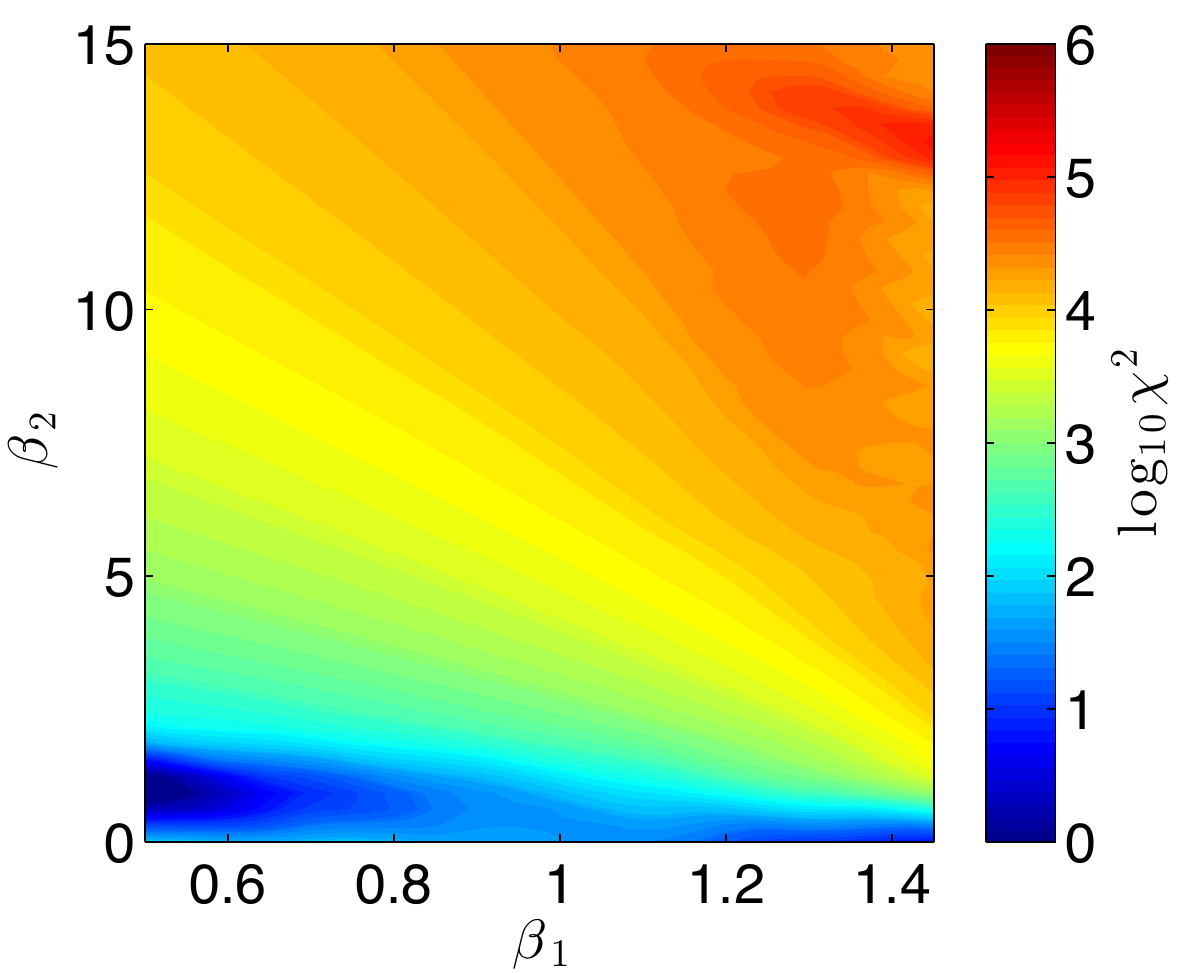}}
		\subfigure[\, Cosmic variance]{\includegraphics[scale=0.5,angle=0]{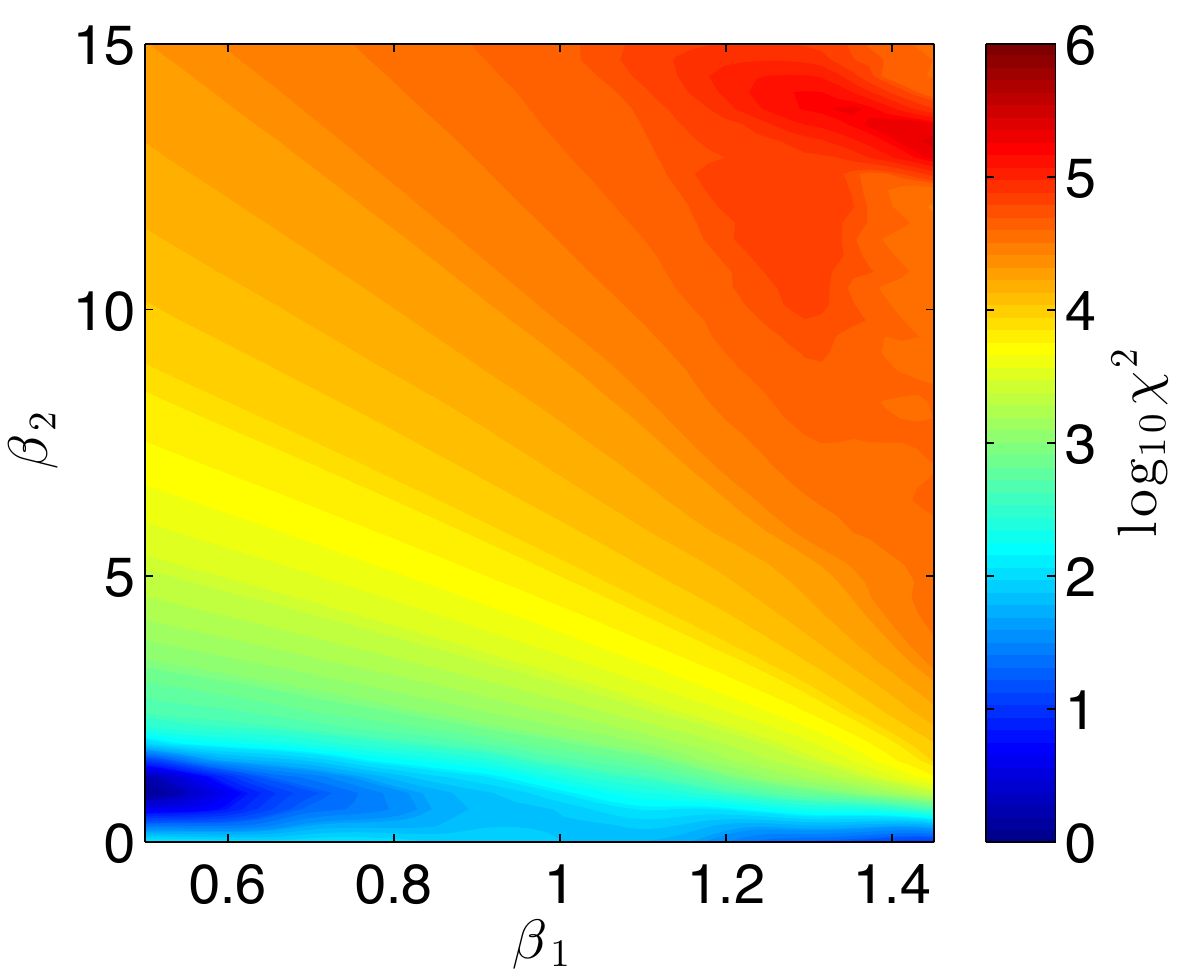}}\\
		\subfigure[\, {\it Euclid}, $\qsubrm{\ell}{max} = 10^2$]{\includegraphics[scale=0.5,angle=0]{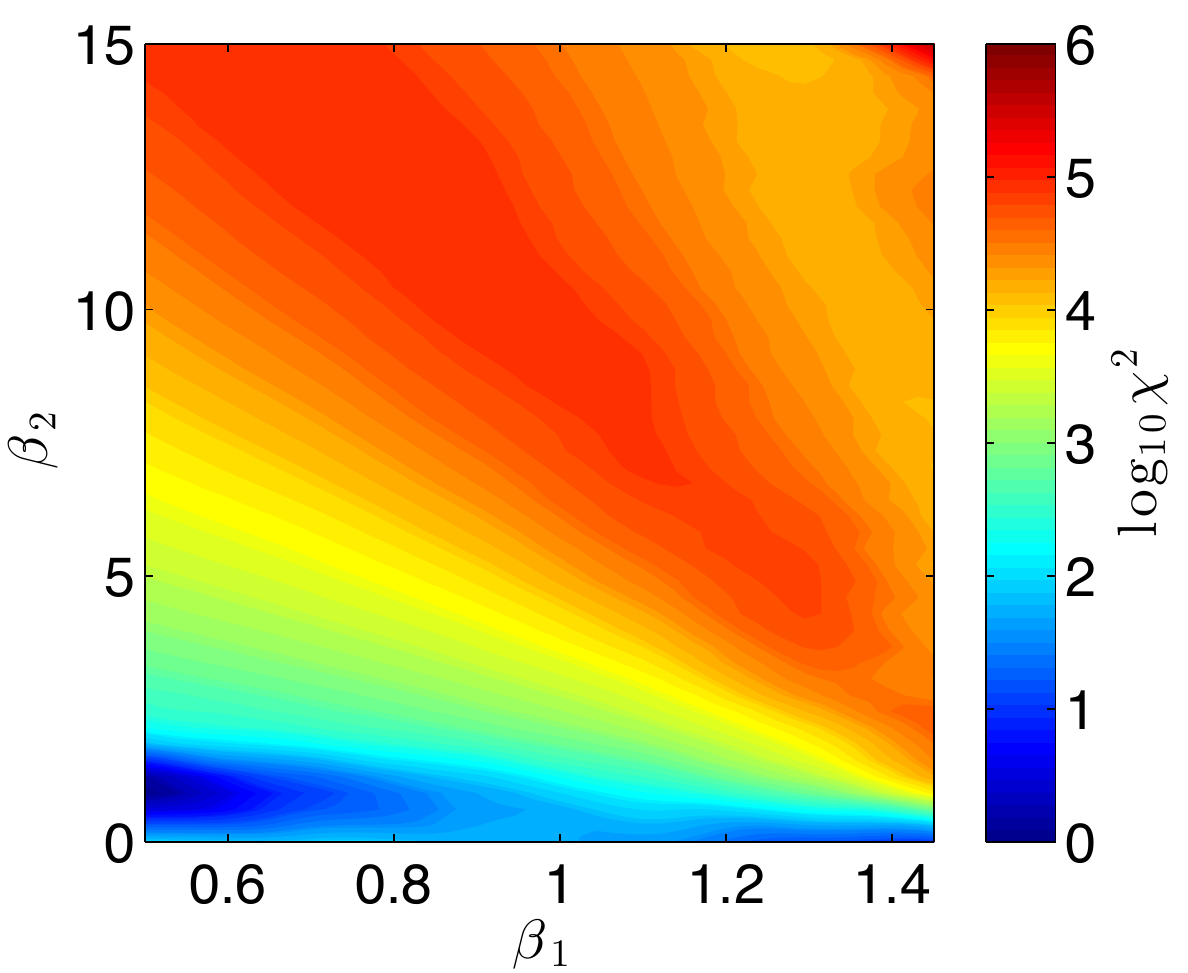}}
		\subfigure[\, {\it Euclid}, $\qsubrm{\ell}{max} = 10^3$]{\includegraphics[scale=0.5,angle=0]{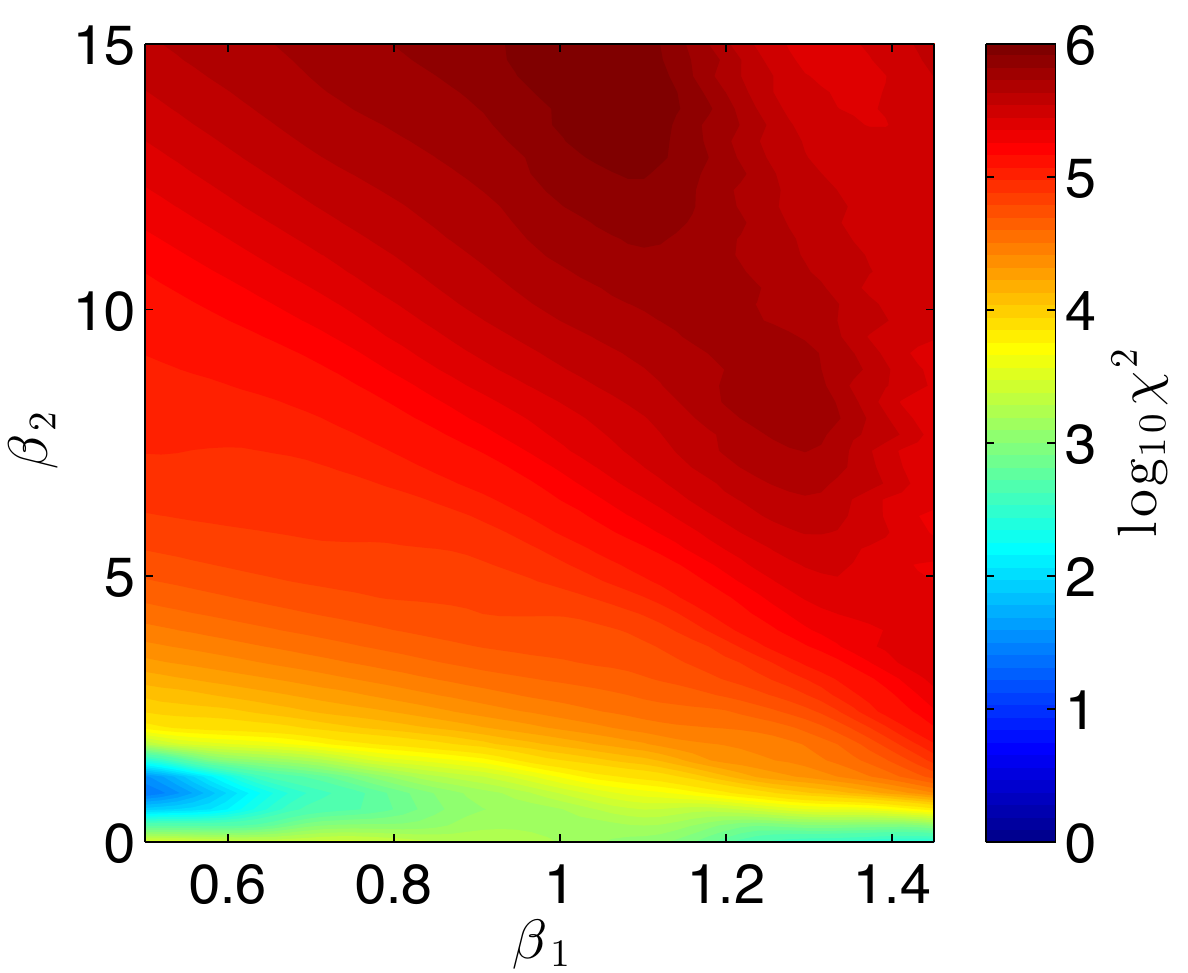}}
      	\end{center}
\caption{Potential discriminatory power of the GSF models; here we have fixed $\alpha = 10^{-5}$ and varied $\beta_1$ and $\beta_2$ as shown in the axes. In (a) and (b) we use CMB experiments -- with PRISM and cosmic variance limited noises, and in (c) and (d) we use tomographic weak lensing experiments -- with \textit{Euclid} noises and with a maximum multipole $\qsubrm{\ell}{max} = 10^2$ and $\qsubrm{\ell}{max}=10^3$ respectively.  }\label{fig:chisq_gen-GSF}

\end{figure}

 In \fref{fig:chisq_gen-GSF} we plot the $\chi^2$ values for various GSF models -- in the plots we fixed the value $\alpha = 10^{-5}$, and varied $\beta_1$ and $\beta_2$. We can see that the constraining power of PRISM compared to the ``ultimate limit'' of cosmic variance limited CMB experiments are very similar. There will always be a degeneracy in measuring $\beta_1$ for low values of $\beta_2$. In the tomographic weak lensing case, increasing $\qsubrm{\ell}{max}$ from $10^2$ to $10^3$ breaks this degeneracy in $\beta_1$ for low values of $\beta_2$.
 
\section{Conclusions}
\label{section:conclusions}
In this paper we presented constraints from current data on the equations of state for perturbations in two distinct classes of models: the GSF   and TDI $\ld(g)$ models. In addition we have explored how well future experiments will be able to constrain the parameters and how further understanding non-linearities will help break observational degeneracies. We presented analytic arguments explaining how the lensing potential should be expected to look rather different in models with and without anisotropic stress.

Our constraints using the {\it Planck}+WP+ BAO+CFHTLenS data in  the TDI $\ld(g)$ model appears to suggest that $\qsubrm{c}{s}^2>10^{-4}$ which implies that the Jeans length of the effective dark sector fluid is $>30h^{-1}\,{\rm Mpc}$. This can also constitute a constraint on the masses of gravitons in the corresponding Lorentz violating massive gravity scenario \cite{BattyePearson_connections}. There the graviton masses $m_i^2$ are given in terms of $w$, $\qsubrm{c}{s}^2$ 
\bse
\bea
m_0^2 &=& - M^2,\\
m_1^2 &=& - \tfrac{1}{4}wM^2,\\
m_2^2 &=&- \tfrac{1}{2}\left[ w+\tfrac{3}{4}(1+w)(\qsubrm{c}{s}^2-w)\right]M^2,\\
m_3^2 &=& \tfrac{1}{4}\left[ w^2 - \tfrac{1}{2}\left( \qsubrm{c}{s}^2-w\right)(1+w)\right]M^2,\\
m_4^2 &=& - \tfrac{1}{2}wM^2,
\eea
\ese
and a mass scale $M^2$ defined by $\Omega_{\rm de}=M^2/(3H_0^2)$. Hence, the constraints on $w$ and $\qsubrm{c}{s}^2$ we gave in \fref{fig:ede_planck_cfhtlens} can  be used to constrain the $m_i^2$.

The results outlined in this paper are encouraging since we are able to constrain parameterizations of dark energy and modified gravity using current linear data. In some sense this success is due to our parameterization being phrased at the level of effective fluid equations which directly govern the evolution of observable quantities, rather than at the level of the action or gravitational field equations -- although there is a correspondance between these approaches.

\section*{Acknowledgements}
JAP  is supported by the STFC Consolidated Grant ST/J000426/1.

\providecommand{\href}[2]{#2}\begingroup\raggedright\endgroup

\end{document}